\documentclass[fleqn,usenatbib]{mnras}
\usepackage[T1]{fontenc}
\usepackage{newtxtext,newtxmath}
\usepackage[utf8]{inputenc}
\usepackage{natbib}
\usepackage{graphicx}
\usepackage{hyperref}
\usepackage{threeparttable}
\usepackage{placeins}
\usepackage[normalem]{ulem}
\usepackage{pdflscape}
\usepackage{amsmath}
\usepackage{orcidlink}
\usepackage{multirow}
\usepackage{xcolor}
\usepackage{soul}

\newcommand{\me}{~$\rm M_{\oplus}$}

\defcitealias{cloutier_characterization_2017}{C17}
\defcitealias{sarkis_carmenes_2018}{S18}
\defcitealias{cloutier_confirmation_2019}{C19}
\defcitealias{artigau_line-by-line_2022}{A22}

\title[K2-18 Revisited]{Revisiting Radial Velocity Measurements of the K2-18 System with the Line-by-Line Framework}

\author[Radica et al.]{
Michael Radica\orcidlink{0000-0002-3328-1203},$^{1}$\thanks{E-mail: michael.radica@umontreal.ca}
Étienne Artigau\orcidlink{0000-0003-3506-5667},$^{1,2}$
David Lafrenière\orcidlink{0000-0002-6780-4252},$^{1}$
Charles Cadieux\orcidlink{0000-0001-9291-5555},$^{1}$
Neil J.~Cook\orcidlink{0000-0003-4166-4121},$^{1}$
\newauthor{René Doyon\orcidlink{0000-0001-5485-4675}$^{1,2}$,
Pedro J.~Amado$^{3}$,
Jos\'e~A.~Caballero\orcidlink{0000-0002-7349-1387}$^{4}$,
Thomas Henning$^{5}$,
Andreas Quirrenbach$^{6}$,}
\newauthor{Ansgar Reiners$^{7}$,
and Ignasi Ribas$^{8,9}$}
\\
$^{1}$Institut de Recherche sur les Exoplanètes and Département de Physique, Université de Montréal, 1375 Avenue Thérèse-Lavoie-Roux,\\ Montréal, QC, H2V 0B3, Canada\\
$^{2}$Observatoire du Mont-Mégantic, Université de Montréal, Montr\'eal, QC H3C 3J7, Canada\\
$^{3}$Instituto de Astrof\'{\i}sica de Andaluc\'{\i}a (IAA-CSIC), Glorieta de la Astronom\'{\i}a s/n, E-18008 Granada, Spain\\
$^{4}$Centro de Astrobiolog\'ia (CSIC-INTA), ESAC, Camino Bajo del Castillo s/n, E-28692 Villanueva de la Ca\~nada, Madrid, Spain\\
$^{5}$Max-Planck-Institut f\"ur Astronomie, K\"onigstuhl 17, D-69117 Heidelberg, Germany\\
$^{6}$Landessternwarte, Zentrum f\"ur Astronomie der Universit\"at Heidelberg, K\"onigstuhl 12, D-69117 Heidelberg, Germany\\
$^{7}$Institut f\"ur Astrophysik und Geophysik, Georg-August-Universit\"at, Friedrich-Hund-Platz 1, D-37077 G\"ottingen, Germany\\
$^{8}$Institut de Ci\`encies de l'Espai (ICE, CSIC), Campus UAB, c/~Can Magrans s/n, E-08193 Bellaterra, Barcelona, Spain\\
$^{9}$Institut d'Estudis Espacials de Catalunya (IEEC), c/ Gran Capit\`a 2-4, E-08034 Barcelona, Spain\\
}

\date{Accepted XXX. Received YYY; in original form ZZZ}

\pubyear{2022}

\begin{document}
\label{firstpage}
\pagerange{\pageref{firstpage}--\pageref{lastpage}}
\maketitle

\begin{abstract}
 The cross-correlation function and template matching techniques have dominated the world of precision radial velocities for many years. Recently, a new technique, named line-by-line, has been developed as an outlier resistant way to efficiently extract radial velocity content from high resolution spectra. We apply this new method to archival HARPS and CARMENES datasets of the K2-18 system. After reprocessing the HARPS dataset with the line-by-line framework, we are able to replicate the findings of previous studies. Furthermore, by splitting the full wavelength range into sub-domains, we were able to identify a systematic chromatic correlation of the radial velocities in the reprocessed CARMENES dataset. After post-processing the radial velocities to remove this correlation, as well as rejecting some outlier nights, we robustly uncover the signal of both K2-18\,b and K2-18\,c, with masses that agree with those found from our analysis of the HARPS dataset. We then combine both the HARPS and CARMENES velocities to refine the parameters of both planets, notably resulting in a revised mass and period for K2-18\,c of $6.99^{+0.96}_{-0.99}$\,\me\ and $9.2072\pm0.0065$\,d, respectively. Our work thoroughly demonstrates the power of the line-by-line technique for the extraction of precision radial velocity information. 
\end{abstract}




\section{Introduction}

Exemplified by its role in the foundational discovery of 51\,Pegasi\,b \citep{mayor_jupiter-mass_1995}, precision radial velocity (pRV) measurements, where one observes the gravitational reflex motion of a host star due to an orbiting planet, is one of the key observational techniques upon which exoplanet science rests. Despite having been overtaken by the transit method in terms of raw numbers of planet detected, thanks to dedicated surveys such as \textit{Kepler} \citep{borucki_kepler_2010} and \textit{TESS} \citep{ricker_transiting_2014}, pRV observations have remained critical, not only for the confirmation of planet candidates, but the understanding of their physical properties. When combined with planetary radius information gained through transit observations, mass estimates provided by pRV surveys allow for estimates of exoplanet bulk densities and thus atmosphere scale heights --- information which is critical for atmospheric characterization \citep{seager_theoretical_2000}. Additionally, the spectroscopic capabilities of pRV instruments have recently started to be used to perform exoplanet atmosphere studies; leveraging the relative Doppler shift of planetary and stellar spectral lines as the planet moves in its orbit to probe the composition and structure of exoplanet atmospheres \citep[e.g.,][]{brogi_signature_2012, brogi_detection_2013, brogi_rotation_2016, birkby_detection_2013, birkby_discovery_2017, guilluy_exoplanet_2019, pelletier_where_2021, boucher_characterizing_2021}. 

Ever since the first exoplanet discovery, RV measurement precision has significantly improved, from $\sim$15\,m/s \citep{mayor_jupiter-mass_1995} to better than 50\,cm/s with the latest state-of-the-art spectrographs \citep{pepe_espresso_2021}. However, recent interest in planetary systems around low-mass stars, as well as atmospheric spectroscopy studies, have motivated a shift in pRV instruments from operating in the optical to near-infrared (NIR) wavelengths \citep{artigau_spirou:_2014, kotani_infrared_2014, quirrenbach_carmenes_2018}. pRV observations in the NIR are significantly more challenging than in the optical due to a combination of factors, perhaps the foremost of which is that telluric absorption is much more prominent in the $Y J H K$ wavebands compared to the optical \citep{artigau_telluric-line_2014}. This imparts a strong telluric signal to all pRV observations in the NIR which generally dwarfs the scientific signal of interest. However, other effects such as emission from OH in Earth's atmosphere \citep{rousselot_night-sky_2000}, as well as detector effects such as persistence \citep{artigau_h4rg_2018} also present substantial challenges. 

\citet[][hereafter A22]{artigau_line-by-line_2022} recently presented a novel method for pRV measurements, uniquely suited to the challenges presented by observations in the NIR as well as optical wavelengths, and demonstrate its effectiveness on two test data sets: publicly available data for Proxima Centauri retrieved from the HARPS data archive, and observations of Barnard's Star from the SPIRou Legacy Survey \citep{donati_spirou_2020}. 

In this article, we apply the unique capabilites of the LBL method to archival HARPS and CARMENES RV data of the K2-18 system to attempt to rectify persistent anomalies in the literature. Our work is laid out as follows: in Section \ref{sec: Overview of LBL} we briefly summarize the key aspects of the LBL pRV technique, then outline the current state of knowledge regarding the K2-18 system in Section \ref{sec: Overview of K218}. Section \ref{sec: HARPS Revisited} presents our reanalysis of the HARPS RV time series presented in \citet{cloutier_characterization_2017} and later extended by \citet{cloutier_confirmation_2019}, and Section \ref{sec: CARMENES Revisited}, a consistent reanalysis of the CARMENES RV time series first published by \citet{sarkis_carmenes_2018}. In section \ref{sec: Joint Reanalysis} we combine all available RV data for the K2-18 system to refine the physical and orbital parameters of the planets, and we summarize and conclude in section \ref{sec: Discussion}. 

\section{Overview of the LBL Technique for pRV Measurements}
\label{sec: Overview of LBL}

The earliest RV datasets were derived through the use of the cross-correlation function (CCF) technique, whereby an observed spectrum is cross-correlated with a mask consisting of a comb of delta functions denoting the locations of stellar lines, with each delta function weighted by the line-depth and local signal-to-noise ratio (S/N). The CCF method yields an average line profile, who's central velocity is the overall RV shift of the observed spectrum, and who's higher-order moments (e.g., the full-width-half-max; FWHM) are used as tracers of stellar activity.

The LBL technique \citepalias{artigau_line-by-line_2022}, is an extension of the \citet{bouchy_fundamental_2001} formalism that uses a projection of the residual between a spectrum and the corresponding stellar template onto the derivative of the spectrum to measure a velocity. In the LBL, the analysis is performed on individual `lines' rather than the spectrum all at once. The motivation for such a procedure is quite simple: pRV spectra, especially in the NIR, are well-known to be effected by telluric absorption and emission as well as other detector effects. Regions affected by spurious structures can be modelled through a mixture model to derive an outlier-resistant velocity.

The LBL framework allows for a per-band RV measurement of the velocity, which provides a consistency check of pRV measurement. This is not unique to the LBL, provided that the domain is wide enough, this can also be done in the framework of CCF measurements \citep[e.g.][]{2022arXiv220905814K} or template matching. The subdivision of a dataset into multiple bands enables  a more robust discrimination between signals of a planetary nature, which should be present in periodograms of all bands, versus those originating
from stellar activity, which are known to be wavelength dependent \citep{reiners_detecting_2010}.

\section{The K2-18 Planetary System}
\label{sec: Overview of K218}

The K2-18 system has attracted considerable interest from the exoplanet community in recent years. A habitable-zone mini-Neptune, K2-18\,b was detected around the M2.5V host star \citep{benneke_spitzer_2017} by \citet{montet_stellar_2015} using two transits from K2 photometry. \citet{benneke_spitzer_2017} later confirmed this discovery via the observation of a single transit with \textit{Spitzer}/IRAC 4.5\,µm photometry. Subsequent transmission spectra taken with the Wide Field Camera 3 instrument of the Hubble Space Telescope then famously yielded a detection of water vapour in its atmosphere \citep{benneke_water_2019, tsiaras_water_2019} --- the first such detection for a habitable-zone planet. Although, further studies \citep[e.g.,][]{barclay_stellar_2021, bezard_methane_2022} have claimed that the inferred signature of atmospheric water vapour may be erroneous, and instead due to inhomogeneities in the stellar surface, or indeed methane absorption.

The temperate conditions of K2-18\,b and the detection of water vapour in its atmosphere have made it the subject of much interest. In-depth modelling efforts have detected the hallmarks of disquilibrium chemistry \citep{hu_photochemistry_2021, blain_1d_2021}, and it was also recently proposed that K2-18\,b may be a so-called Hycean world --- an ocean planet surrounded by a thin H/He dominated atmosphere, and potentially even habitable \citep{madhusudhan_interior_2020, madhusudhan_habitability_2021, piette_temperature_2020}. \citet{hu_unveiling_2021} suggest that the unique chemical signatures of such Hycean worlds will even be detectable with the James Webb Space Telescope (JWST). Indeed, to this end K2-18\,b will be targeted for observation during JWST Cycle 1 with the NIRSpec, MIRI, and NIRISS instruments to attempt to shed light on its composition and internal structure\footnote{GO Programs \#2372 \& \#2722}. 

There is more intrigue though, in the K2-18 system than just the nature of K2-18\,b. The first RV analysis on this system was carried out by \citet[][hereafter C17]{cloutier_characterization_2017} who observed 75 spectra of K2-18 from April 2015 to May 2017 with the HARPS spectrograph \citep{pepe_harps_2004}. Not only did these observations provide an independent confirmation of the planetary nature of K2-18\,b, but \citetalias{cloutier_characterization_2017} also claimed the detection of a second, non-transiting planet, K2-18\,c. \citetalias{cloutier_characterization_2017} find a mass of $8.0\pm1.0$\me\ for K2-18\,b, and a minimum mass of $7.5\pm1.3$\me, as well as a period of \hbox{8.96\,d} for K2-18\,c. 

\citet[][hereafter S18]{sarkis_carmenes_2018} presented 58 spectra of K2-18 observed with the VIS channel of the CARMENES spectrograph \citep{quirrenbach_carmenes_2018} between December 2016 and June 2017. They recover a strong signal of K2-18\,b in their RV data, but find no evidence for K2-18\,c. Their analysis finds that the signal of K2-18\,c is only present in the second half of their data set, when the star displayed an increased level of activity (as demonstrated by the level of signal from the Ca infrared triplet; see their Fig.~8). They thus conclude that the $\sim$9-day signal found by \citetalias{cloutier_characterization_2017} is likely do to stellar activity. 

\citet[][hereafter C19]{cloutier_confirmation_2019} then revisited their initial analysis in conjunction with the CARMENES observations presented by \citetalias{sarkis_carmenes_2018} as well as 31 additional HARPS spectra - extending the full K2-18 baseline from April 2015 to June 2018. Through a thorough reanalysis of the CARMENES measurements, they conclude that three `anomalous' nights are artificially suppressing the signal of K2-18\,c, and that once removed, the signal is much more apparent in a Lomb-Scargle periodogram \citep{scargle_studies_1982} analysis. They then proceed to a joint analysis of all available RV data for K2-18 (minus the three anomalous nights), as well as individual analyses of just the HARPS and CARMENES RVs separately. The parameters of K2-18\,b are consistent across both sets of data, and the joint analysis refines its mass to $8.63 \pm 1.35$\me. However, HARPS prefers a significantly larger median RV semi-amplitude, and therefore larger minimum mass, for K2-18\,c than does CARMENES (although the two are consistent at $1\sigma$; see their Fig.~6). The joint analysis thus favours a minimum mass intermediate to that calculated from either HARPS or CARMENES alone, of $5.62 \pm 0.84$\me --- $\sim$ 2$\sigma$ lower than the original estimate from \citetalias{cloutier_characterization_2017}, as well as a period of 8.99\,d --- a 4.3$\sigma$ difference from their original estimate. A summary of the most pertinent planet parameters from each study (including this present one) is included in Table~\ref{tab: Comparison}.

\begin{table*}
\caption{Comparison of Planet Parameters from Different Studies}
\label{tab: Comparison}
\begin{threeparttable}
    \begin{tabular}{ccccc}
        \hline
        \hline
        Parameter & \multicolumn{4}{c}{Study} \\
         & \citetalias{cloutier_characterization_2017} & \citetalias{sarkis_carmenes_2018} & \citetalias{cloutier_confirmation_2019} & This Study \\
        \hline

        \textit{K2-18\,b} & & & & \\
        Period, $P_b$ [days] & $32.9396\pm10^{-4}$ & $32.9396\pm10^{-4}$ & $32.9396\pm10^{-4}$ & $32.9396\pm10^{-4}$\\
        RV Semi-Amplitude, $K_b$ [m/s] & $3.18\pm0.71$ & $3.55\pm0.57$ & $2.75\pm0.43$ & $3.112\pm0.56$ \\

        \textit{K2-18\,c} & & & & \\
        Period, $P_c$ [days] & $8.962\pm0.008$ & -- & $8.997\pm0.007$ & $9.2072\pm0.0065$ \\
        RV Semi-Amplitude, $K_c$ [m/s] & $4.63\pm0.72$ & -- & $2.76\pm0.41$ & $3.568\pm0.45$ \\
        \hline
        
    \end{tabular}
    \begin{tablenotes}
        \small
        \item \textbf{Notes}: -- denotes that insufficient evidence for the planet was found in the study. 
    \end{tablenotes}
\end{threeparttable}
\end{table*}

\section{Analysis and Results}
\subsection{HARPS Revisited}
\label{sec: HARPS Revisited}
We retrieved all available processed HARPS spectra for K2-18 from the ESO science archive\footnote{\url{http://archive.eso.org/wdb/wdb/adp/phase3_main/form}}. A total of 100 nights were retrieved covering a time period of April 2015 to July 2018. The first 75 nights of these data, up to May 2017, were previous independently reduced and analyzed in \citetalias{cloutier_characterization_2017}, and \citetalias{cloutier_confirmation_2019} later presented the latter 25 nights of observations. We then passed all 100 nights through the LBL pipeline to extract the precision radial velocity information. 

\begin{figure}
	\centering
	\includegraphics[width=\columnwidth]{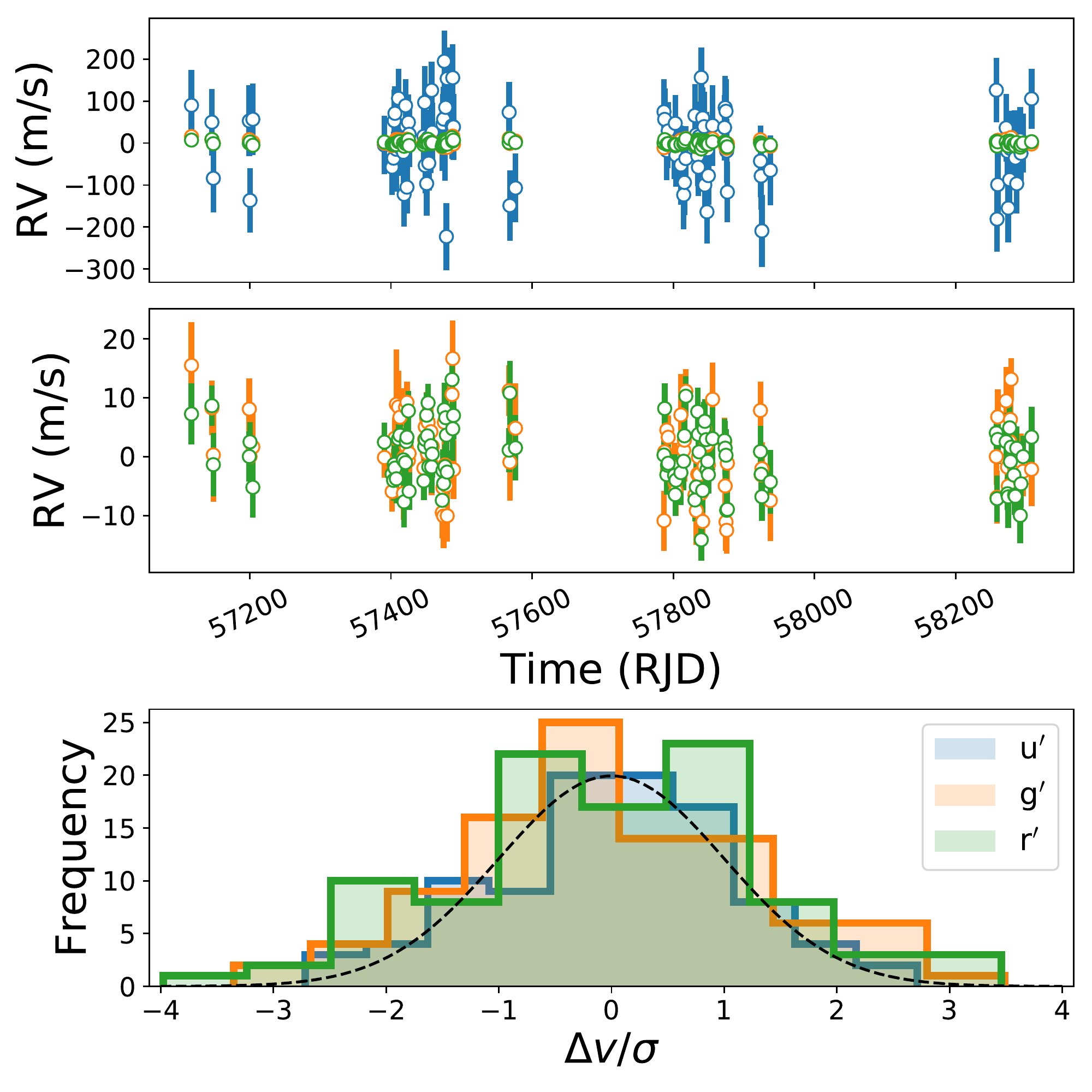}
    \caption{\emph{Top}: Median subtracted RV values for each of our three chosen HARPS bands.
    \emph{Middle}: Same as the top, but with the $u^\prime$ band removed to better visualize the other two bands.
    \emph{Bottom}: Histogram showing the velocity difference from the mean, divided by the RV error for each band. The distribution for each band traces a normal distribution (dashed black curve), and no $>$5$\sigma$ outliers are present.  
    \label{fig:HARPS_hist}}
\end{figure}

To verify self-consistency, we subdivide the full HARPS wavelength range (378 -- 691\,nm) into three bands, corresponding to the $u^\prime$ (324 -- 391\,nm), $g^\prime$ (337 -- 613\,nm), and $r^\prime$ (496 -- 744\,nm) bands of the Sloan Digital Sky Survey (SDSS) \citep{fukugita_sloan_1996}. We then perform an iterative sigma clip on each band to remove 3$\sigma$ outliers in RV and RV error. Three nights are removed in this way (the same three nights were flagged in each band, and were thus removed). The resulting RV values are shown in Fig.~\ref{fig:HARPS_hist}. The \citet{bouchy_fundamental_2001} framework should lead to normally-distributed errors, and we first verify this by constructing histograms of the per-night velocity difference from the mean, divided by the corresponding RV uncertainty for each night. As shown in Fig.~\ref{fig:HARPS_hist}, all three bands nicely appear trace a normal distribution, with no remaining $>$5$\sigma$ outliers. We verify the Gaussian nature of these distributions using the D'Agostino K-squared test \citep{dagostino_tests_1973} via the \texttt{normaltest} routine of the \texttt{scipy.stats} package, and indeed find $p$-values for each band of $>$0.85, indicating consistency with a normal distribution in each case. The statistics for each band are provided in Table~\ref{tab: Band Comparison}.

\begin{table}
    \caption{Comparison of Individual Bands for the LBL Reprocessed CARMENES and HARPS Datasets}
    \label{tab: Band Comparison}
    \begin{tabular}{cccc}
        \hline
        \hline
        \multirow{2}{*}{Band} & \multirow{2}{*}{RMS [m/s]} &\multirow{2}{*}{RMS Error [m/s]} & Fractional \\
        & & & Contribution [\%]\\
        \hline

        HARPS & & & \\
        $u^\prime$ & 82.53 & 75.42 & 0.17\\
        $g^\prime$ & 6.15 & 4.90 & 42.51\\
        $r^\prime$ & 5.17 & 4.11 & 57.32\\
        \vspace{2mm}Total & 4.82 & 2.73 &  \\
        
        CARMENES & & & \\
        $g^\prime$ & 13.02 & 9.03 & 7.25\\
        $r^\prime$ & 8.34 & 2.54 & 84.01\\
        $i^\prime$ & 12.06 & 8.14 & 8.06\\
        $z^\prime$ & 41.11 & 27.51 & 0.69\\
        Total & 8.25 & 2.33 &  \\
        \hline
        
    \end{tabular}
\end{table}

\begin{figure}
	\centering
	\includegraphics[width=1\columnwidth]{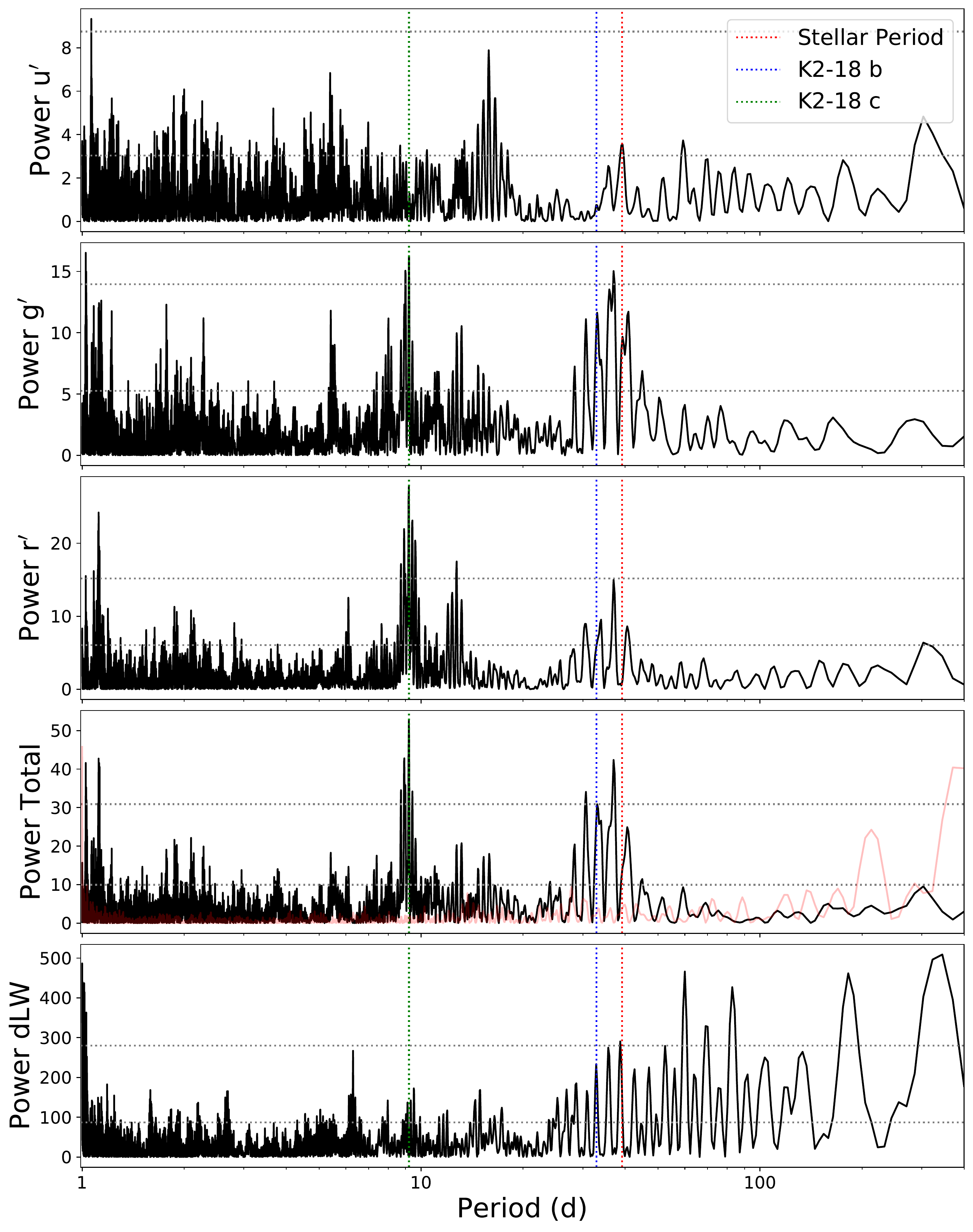}
    \caption{Lomb-Scargle periodograms for each of our three HARPS bands, as well as the total RV and dLW parameter. The derived orbital periods of K2-18\,b, and c, as well as the stellar rotation period (from Section \ref{sec: HARPS Revisited}) are denoted with blue, green, and red dotted vertical lines respectively. The 1\%, and 0.1\% false alarm probabilities are denoted via the grey horizontal dotted lines. The next to last panel also displays the periodogram of the window function in red.  
    \label{fig:HARPS_periodogram}}
\end{figure}

We proceed to construct Lomb-Scargle periodograms \citep{scargle_studies_1982} for each band, as well as for the total nightly RV measurement (which we construct as the average of the RV measurement in each band weighted by the corresponding RV error), which we show in Fig.~\ref{fig:HARPS_periodogram}. The total RV periodogram is qualitatively similar to that which was presented in \citetalias{cloutier_confirmation_2019} (see their Fig.~A.2.). Strong signals are present at the predicted orbital periods of K2-18\,b, and c, as well as the stellar rotation period. Similar signals are also present in the $g^\prime$, and $r^\prime$ bands, although the signal of the $\sim$32\,day stellar rotation period is less significant in the $r^\prime$ band than the $g^\prime$. There are no significant signals present in the $u^\prime$ band, which is unsurprising given the large error bars and scatter in this band (e.g., Fig.~\ref{fig:HARPS_hist}). We could cut the entirety of the $u^\prime$ band, and proceed with using only the $g^\prime$ and $r^\prime$ bands. However, given that the distribution of the $u^\prime$ band velocities are still Gaussian, and that, due to the large error bars the $u^\prime$ band contributes at an extremely low level ($\sim$1\%) to the combined RV signal, we elect to retain it for our analysis. We note though, that even if we do cut the $u^\prime$ band entirely, our results remain unchanged.

\begin{figure}
	\centering
	\includegraphics[width=1.1\columnwidth]{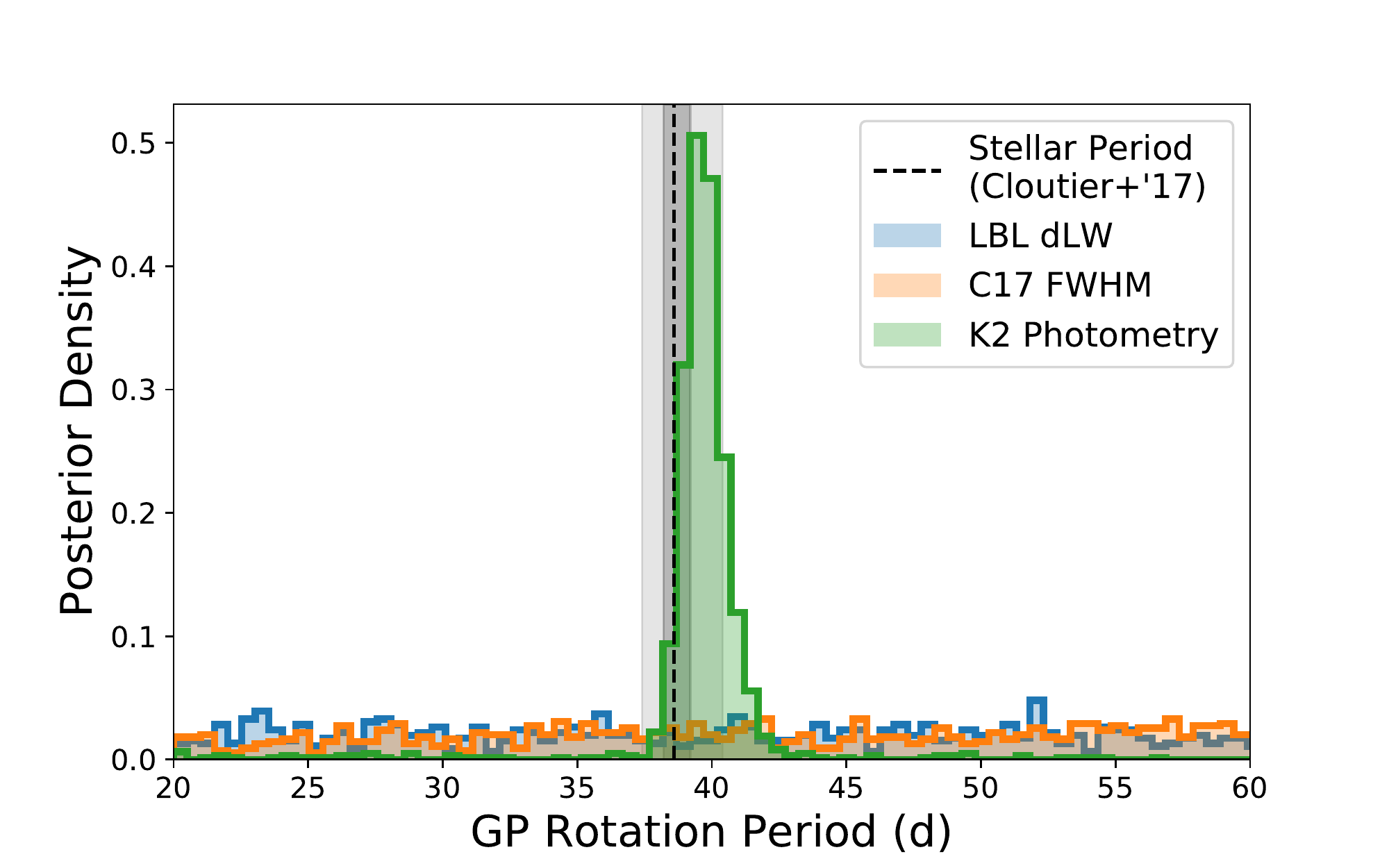}
    \caption{Posterior distributions for the $P_{rot}$ parameter of the exponential-sine-squared GP (Equation~\ref{equ: SSE kernel}) fit with \texttt{juliet}, resulting from training on three different datasets: the LBL dLW, the FWHM values published by \citetalias{cloutier_characterization_2017}, and the K2 photometry. Only the the K2 photometry training set yields a strong constraint on $P_{rot}$. A uniform prior from zero to 100 days was used in all cases. 
    \label{fig:prot_posteriors}}
\end{figure}

Following \citetalias{cloutier_characterization_2017, cloutier_confirmation_2019}, we then proceed to model the total RV measurements. Common practice in the RV literature is to employ Gaussian Processes (GPs) within a Bayesian retrieval framework to efficiently model stellar activity signals \citep[e.g.,][]{gilbertson_toward_2020}. To fully leverage the capabilities of a GP, it is important to `train' a GP on ancillary data --- in the scope of RV analysis where the goal is to model stellar activity, training sets include common activity indicators such as the H$\alpha$ index, as well as the full width at half-maximum (FWHM) or bi-sector inverse slope of the cross-correlation function constructed during RV extraction. \citetalias{artigau_line-by-line_2022} demonstrated that the LBL can calculate the change in line width parameter, dLW \citep{zechmeister_spectrum_2018}. The dLW is linked to the change in the FWHM of lines, and is equal to the FWHM for Gaussian line profiles. The dLW periodicity is shown in the bottom panel of Fig.~\ref{fig:HARPS_periodogram}, and there is some power at the stellar rotation period. However, we find that training a GP on the LBL dLW does not result in any meaningful constraint on the stellar rotation period. Training on the FWHM values published by \citetalias{cloutier_characterization_2017} yields an identical result --- indicating that the use of either of these activity indicators as training sets would not add meaningful constraints to the GP model. Training on the K2 photometry though, does result in a strong constraint on the stellar rotation period of $39.669\pm0.809$\,d, which is in agreement with that derived with the same method by \citetalias{cloutier_characterization_2017}, as well as that from \citetalias{sarkis_carmenes_2018}. Therefore, similarly to both \citetalias{cloutier_characterization_2017} and \citetalias{cloutier_confirmation_2019}, we choose to retain the K2 photometry as our training data. The stellar rotation period posterior distributions for each of these three training sets are shown in Figure~\ref{fig:prot_posteriors}.

 K2-18 was observed during Campaign 1 by the K2 mission between June and August 2014. As mentioned above, to retain consistency with \citetalias{cloutier_characterization_2017} and  \citetalias{cloutier_confirmation_2019}, we train our GP model on the K2 photometry. To this end, we obtained the full, detrended photometric time series from the MAST archive\footnote{\url{http://archive.stsci.edu/k2/hlsp/everest/search.php}}. Again, following \citetalias{cloutier_characterization_2017} we selected the EVEREST reduction \citep{luger_everest_2016}, and removed the two observed transits of K2-18\,b in order to create our training set. We note though, that we also tested other reductions of the K2 lightcurves including POLAR \citep{barros_new_2016} and K2SFF \citep{2014PASP..126..948V} --- the results in each case are completely consistant. 

We jointly fit the training photometry and RV data with the \texttt{juliet} package \citep{espinoza_juliet_2019}. For each planet we fit a standard five-parameter Keplerian orbit ($P, t_0, \sqrt{e}\sin\omega, \sqrt{e}\cos\omega, K$). To both the RV and photometry, we fit an exponential-sine-squared GP as implemented by the \texttt{george} package \citep{ambikasaran_fast_2016}, and built into \texttt{juliet}. Exponential-sine-squared GP models are inherently periodic, and have had much success in modelling stellar activity. \texttt{george} defines the exponential-sine-squared kernel $k$, on some data $x$, as:

\begin{equation}
    k(x) = \sigma^2 \exp\bigg(-\alpha x^2 - \Gamma\sin^2\bigg[\frac{\pi x}{P_{rot}}\bigg]\bigg) ,
    \label{equ: SSE kernel}
\end{equation}

\noindent via four hyperparameters, $\sigma, \alpha, \Gamma, P_{\rm rot}$. The $\sigma$ parameter governs the amplitude of the GP, and thus the amplitude of the stellar variations. We thus fit $\sigma$ separately for the photometry and RVs, whereas the other three parameters govern the length scales of the exponential and sinusoidal variations, and are thus shared between the two data sets. We additionally fit an additive scalar jitter term individually to both datasets to account for potential under-estimations of the error bars, as well as for the systematic velocity of the K2-18 system. In total, our fit has 18 free parameters, the assumed prior distributions for which are listed in Table~\ref{tab: JOINT Parameters}.

\begin{figure*}
	\centering
	\includegraphics[width=\textwidth]{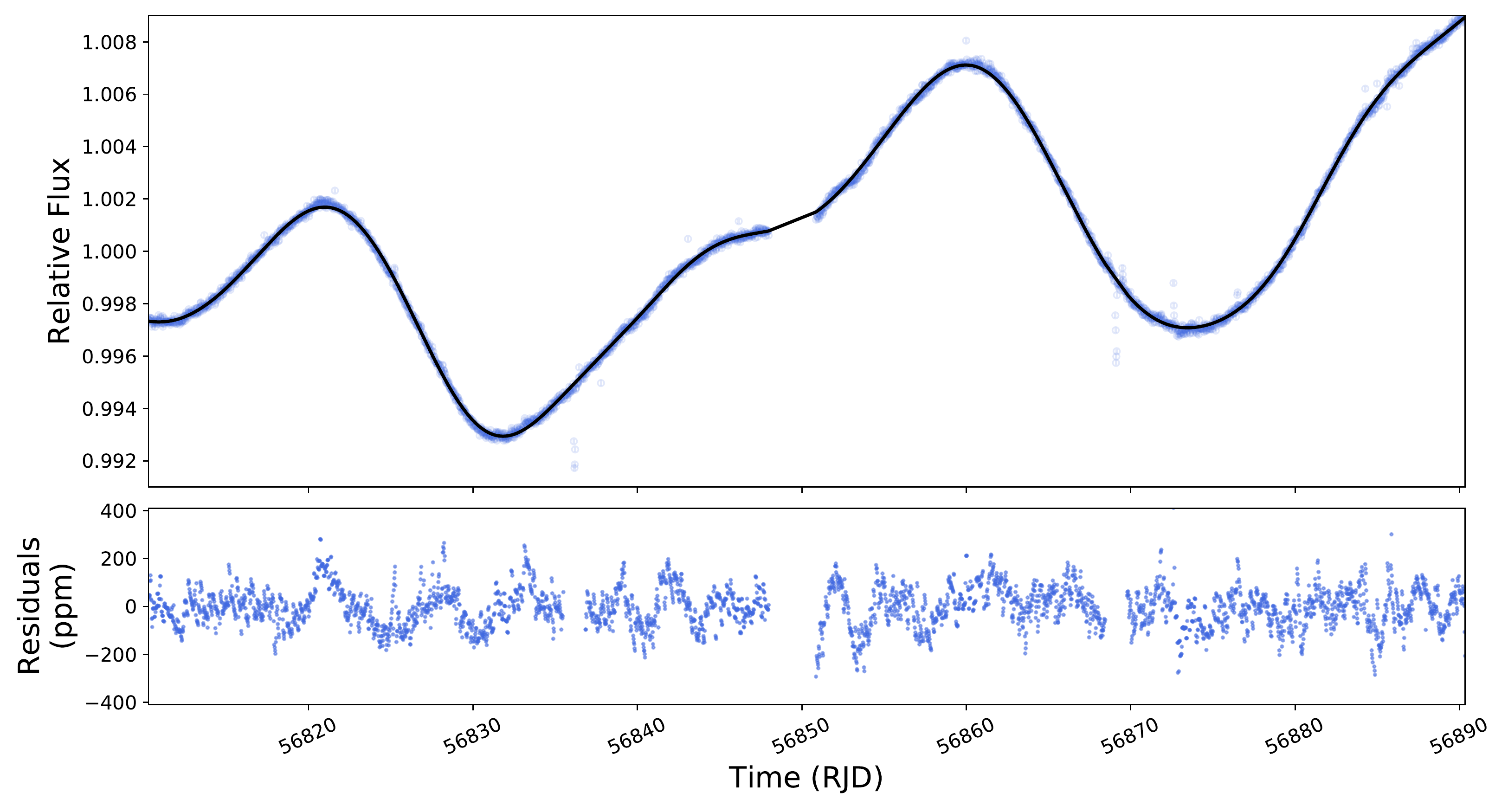}
    \caption{\emph{Top}: K2 photometry (blue points) of K2-18 showing quasi-periodic modulation indicative of stellar activity. The best-fitting exponential-sine-squared GP model is overplotted in black.
    \emph{Bottom}: Residuals to the light curve fit.
    \label{fig:JOINT Photometry}}
\end{figure*}

\texttt{juliet} implements sampling via both Markov-Chain Monte Carlo (MCMC) and Nested Sampling algorithms. We use Nested Sampling through the \texttt{MultiNest} algorithm \citep{feroz_multinest_2009}, which is implemented in \texttt{juliet} via \texttt{PyMultiNest} \citep{buchner_statistical_2016}. Nested Sampling has numerous benefits over MCMC, including the ability to better map multi-modal posterior distributions, as well as directly calculating the Bayesian evidence, $Z$, which enables model comparison \citep{skilling_nested_2006}. The fit results to the photometry are shown in Fig.~\ref{fig:JOINT Photometry}, and to the RVs on the left side of Fig.~\ref{fig:JOINT RVModel}. The posterior distributions are shown in blue in Fig.~\ref{fig:JOINT Corner} as well as listed in Table~\ref{tab: JOINT Parameters}.

Direct comparisons between our results and either \citetalias{cloutier_characterization_2017} or \citetalias{cloutier_confirmation_2019} are difficult as \citetalias{cloutier_characterization_2017} only analyzed the first 75 nights of data, whereas \citetalias{cloutier_confirmation_2019} only report the results of their joint analysis with the CARMENES data. Nevertheless, the majority of our results are consistent with the findings of both studies. Comparing our derived RV semi-amplitudes for both planets to the HARPS-only results presented in Fig.~6 of \citetalias{cloutier_confirmation_2019}, we find that our results are consistent at a 1$\sigma$ level, and the marginalized posterior distributions have qualitatively similar widths, indicating that the precision on the RV semi-amplitudes derived from LBL RV data is comparable to those derived using template-matching. This result is not surprising, as both methodologies are based upon the \citet{bouchy_fundamental_2001} framework, and should thus yield consistent results. 

The only major discrepancy between our reanalysis, and the results published in \citetalias{cloutier_confirmation_2019} is the period of K2-18\,c. \citetalias{cloutier_characterization_2017} derive a period of $8.962\pm0.008$\,d, which is further refined to $8.997\pm0.007$\,d in \citetalias{cloutier_confirmation_2019}. However, our analysis yields a period of $9.207\pm0.006$\,d --- a 35$\sigma$ discrepancy. To further investigate this inconsistency, we apply our fitting procedure to both the ensembles of data presented in \citetalias{cloutier_characterization_2017} and \citetalias{cloutier_confirmation_2019}; that is, the RV measurements extracted with the \texttt{NAIRA} template matching algorithm instead of the LBL. Our reanalysis of the \citetalias{cloutier_characterization_2017} ensemble yields a period which is 1-$\sigma$ consistent with their findings. However, using the full \citetalias{cloutier_confirmation_2019} ensemble, we once again retrieve a 9.207\,day period --- not the 8.997\,d that \citetalias{cloutier_confirmation_2019} found on the exact same dataset. We therefore conclude that it is not the difference in RV extraction routines (LBL versus template matching) which causes this difference.

\citetalias{cloutier_confirmation_2019} use an MCMC algorithm for their RV fits, as opposed to our choice of Nested Sampling. They initialize their walkers at the best-fitting values of each parameter as calculated in \citetalias{cloutier_characterization_2017}. Therefore, if the posterior distribution is sufficiently multi-modal, it is possible for the walkers to get stuck in one mode, and not explore the entire parameter space --- which would lead to them finding another mode centered around 9.21\,d. To test this hypothesis, we switch to \texttt{juliet}'s MCMC sampler, which is implemented through the \texttt{emcee} package \citet{foreman-mackey_emcee_2013} --- the same sampler used by \citetalias{cloutier_confirmation_2019}. We initialize each walker at the best fitting parameters from \citetalias{cloutier_characterization_2017} and fit both the LBL and template matching RV data. In both cases, the MCMC converges on a period of $\sim$8.99\,d, although in the LBL case, some walkers make the jump to the $\sim$9.21\,d mode, revealing the true bi-modal nature of the posterior distribution. Indeed, if we instead initialize the walkers around a period of 9.21\,d for K2-18\,c, both datasets yield the 9.207\,d period. 

We thus conclude that the different periods are due to inefficiencies in the MCMC sampler when exploring multi-modal posteriors and/or an inadequate choice of starting positions for the MCMC walkers --- issues which are not encountered by Nested Sampling routines. To further solidify the validity of our result, we performed two Nested Sampling fits to the LBL data fixing the period of K2-18\,c to 8.997\,d in one case, and 9.207\,d in the other. By comparing the Bayesian evidence values, we find that the 9.207\,d period is strongly preferred by $\Delta \ln Z=5.44$ or $>$3.6$\sigma$ \citep{benneke_how_2013}. Furthermore, K2-18\,c itself is identified in the data at a $\sim$4$\sigma$ level ($\Delta\ln Z=6.49$).

\subsection{CARMENES Revisited}
\label{sec: CARMENES Revisited}
\begin{figure}
	\centering
	\includegraphics[width=\columnwidth]{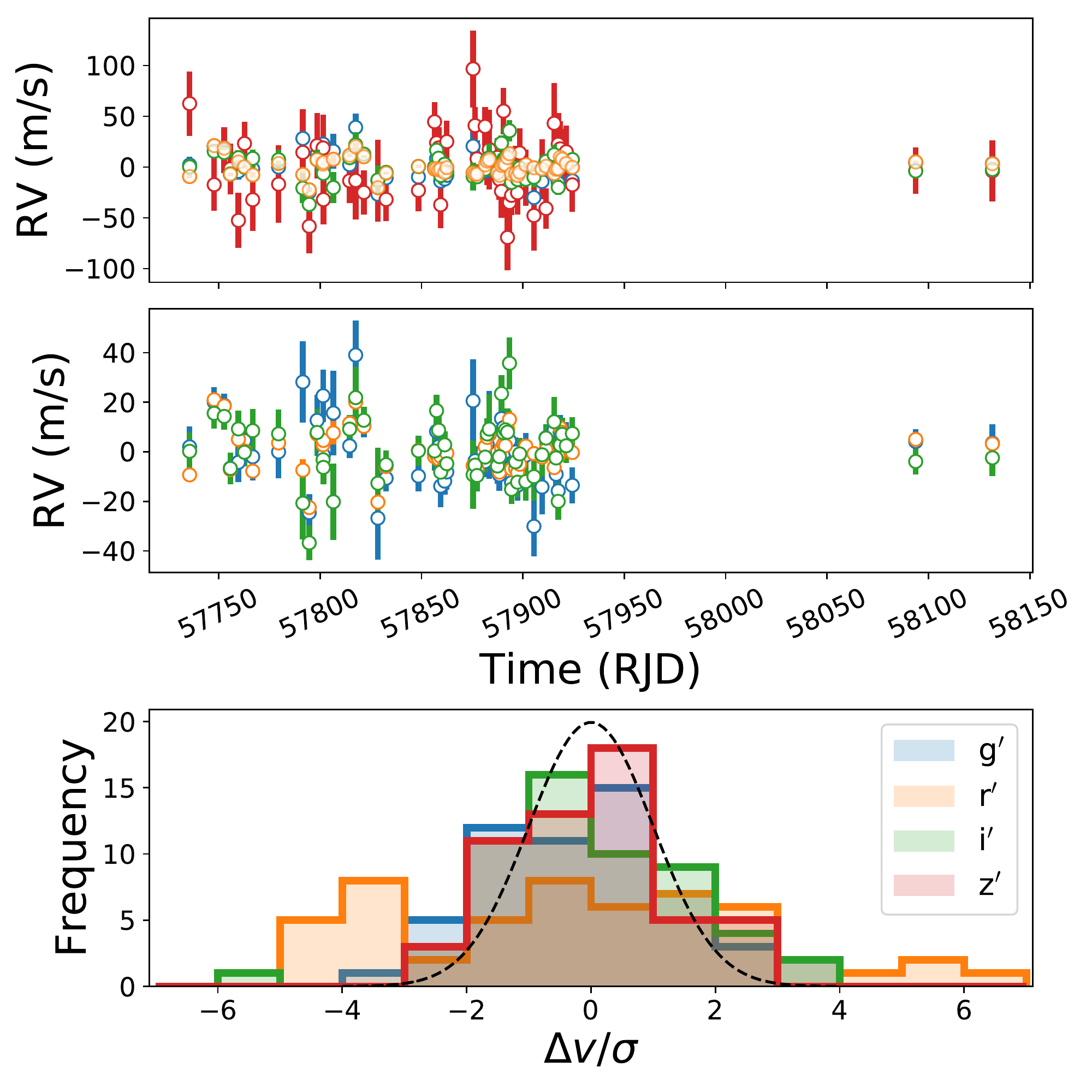}
    \caption{\emph{Top}: Median subtracted RV values for each of our four chosen CARMENES bands.
    \emph{Middle}: Same as top, but with the $z^\prime$ band removed to better visualize the other three bands.
    \emph{Bottom}: Histogram showing the velocity difference from the mean, divided by the RV error for each band. 
    \label{fig:CARMENES_hist}}
\end{figure}

\begin{figure}
	\centering
	\includegraphics[width=1\columnwidth]{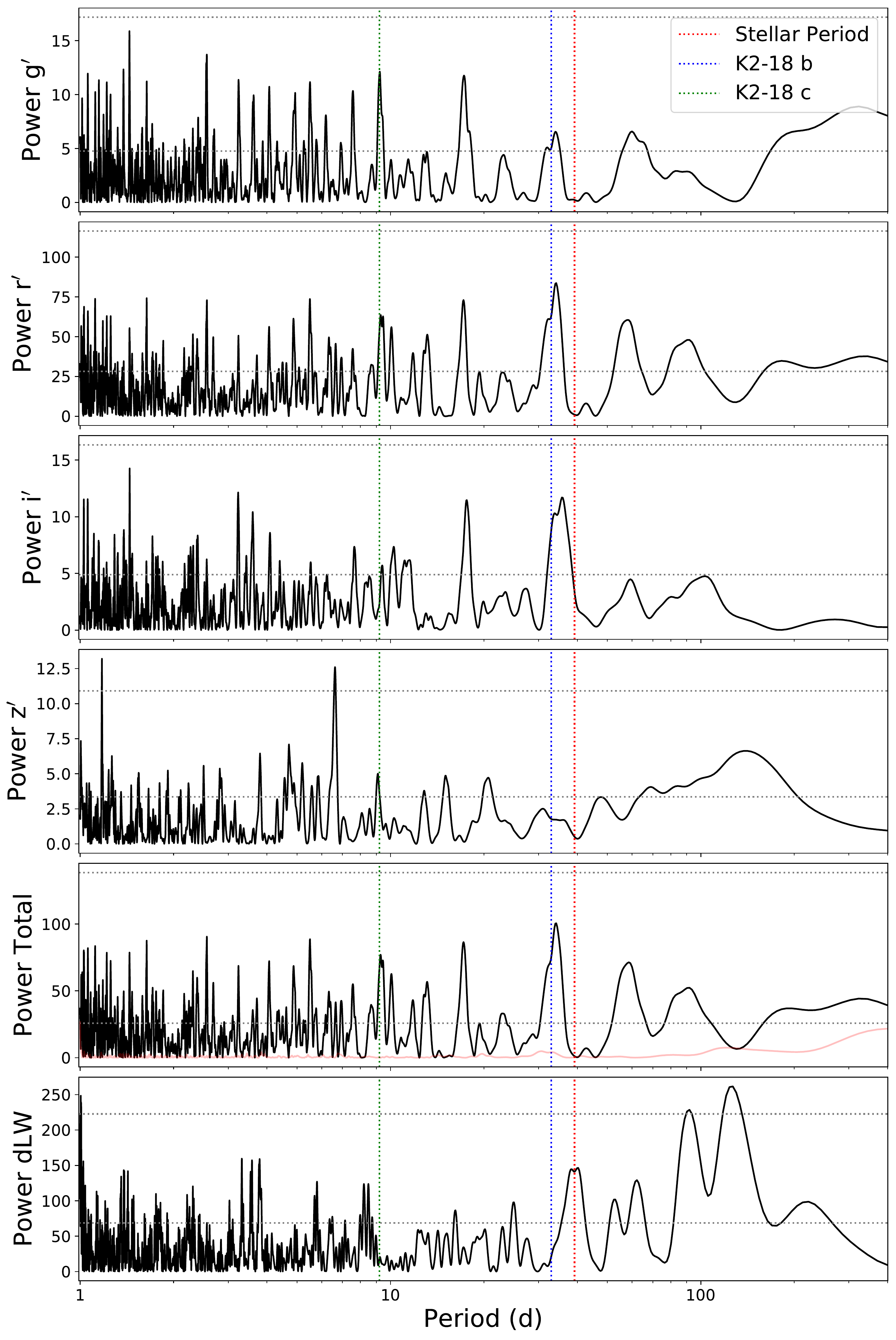}
    \caption{Identical to Fig.~\ref{fig:HARPS_periodogram} but for the four CARMENES bands. 
    \label{fig:CARMENES_periodogram}}
\end{figure}

We next turn our attention to the CARMENES-VIS dataset (520 -- 960\,nm) first published by \citetalias{sarkis_carmenes_2018}. We reprocess the CARMENES spectra through the LBL algorithm in the same way as the HARPS, and divide the full CARMENES bandpass into four individual bands, corresponding roughly to the SDSS $g^\prime$, $r^\prime$, $i^\prime$, and $z^\prime$ (711 -- 1221\,nm) bands. We then perform initial processing (3$\sigma$ outliers in RV and RV error) and visualizations in the same manner as for the HARPS data. Our full CARMENES dataset includes 64 nights of data spanning the time period from December 2016 to February 2018. Only 58 nights were presented in \citetalias{sarkis_carmenes_2018}: our dataset contains three nights of data taken after June 2016 (which was the latest date included in \citetalias{sarkis_carmenes_2018}), as well as three additional nights observed prior to June 2016. The RV time series for each of the four bands, as well as the consistency histograms are shown in Fig.~\ref{fig:CARMENES_hist}, and the periodograms are shown in Fig.~\ref{fig:CARMENES_periodogram}. In general, our reprocessed CARMENES measurements maintain a slightly higher precision than those first presented by \citetalias{sarkis_carmenes_2018} (RMS error of 2.33 vs 3.60~m/s) but a larger overall scatter (RMS of 8.25 vs 5.78~m/s). The full statistics for each band are shown in Table~\ref{tab: Band Comparison}.

Unlike for the HARPS dataset, the histograms do not all trace a normal distribution. Indeed, when the D'Agostino K-squared test is applied, each case yields a $p$-value $<$10$^{-9}$, indicating strong deviation from normality. This becomes especially concerning when inspecting the periodograms for each band. There are no signals higher than the 0.1\% FAP line near the expected periods of either planet, nor the expected stellar rotation period in any of the CARMENES periodograms. Signals around the orbital periods of both planets, as well as the rotation period of the star are prominent, but barely above the level of the ``noise'' at shorter ($<$5\,d periods). Our periodograms are comparatively noisier than those presented by \citetalias{sarkis_carmenes_2018}, although the roots of this additional noise are not clear. Comparing the total and $r^\prime$ band periodograms, it appears that two are nearly identical. This is not surprising, given that the precision obtainable in the $r^\prime$ band is much greater than what is possible in the other three bands. It therefore contributes $\sim$85\% of the total RV signal --- making the non-Gaussian nature of its histogram all the more concerning. 

We attempt a first fit, identical to the fits we performed with the HARPS data --- fitting the same ensemble of parameters jointly to the CARMENES RV data and the K2 photometry. We are able to recover the signal of K2-18\,b with a RV semi-amplitude consistent with that found in the HARPS analysis, but only at a $\sim$2$\sigma$ level, and only an upper limit is retrieved for K2-18\,c. We experiment with removing the $r^\prime$ band entirely, and only keeping the three other bands whose histograms are roughly Gaussian. However, in this case, we still only retrieve an upper limit for K2-18\,c, but also lose the signal of K2-18\,b. This is unsurprising, given how strongly the $r^\prime$ band contributes to the total RV signal, that removing it removes much of the scientific signal of interest. 

We had hoped that the unique capabilities of the LBL methodology would shed light on the reason why the signal of K2-18\,c appear to be suppressed in the CARMENES dataset. We therefore set out to uncover the root causes of the non-Gaussian nature of the $r^\prime$ band histogram.

\begin{figure}
	\centering
	\includegraphics[width=1.1\columnwidth]{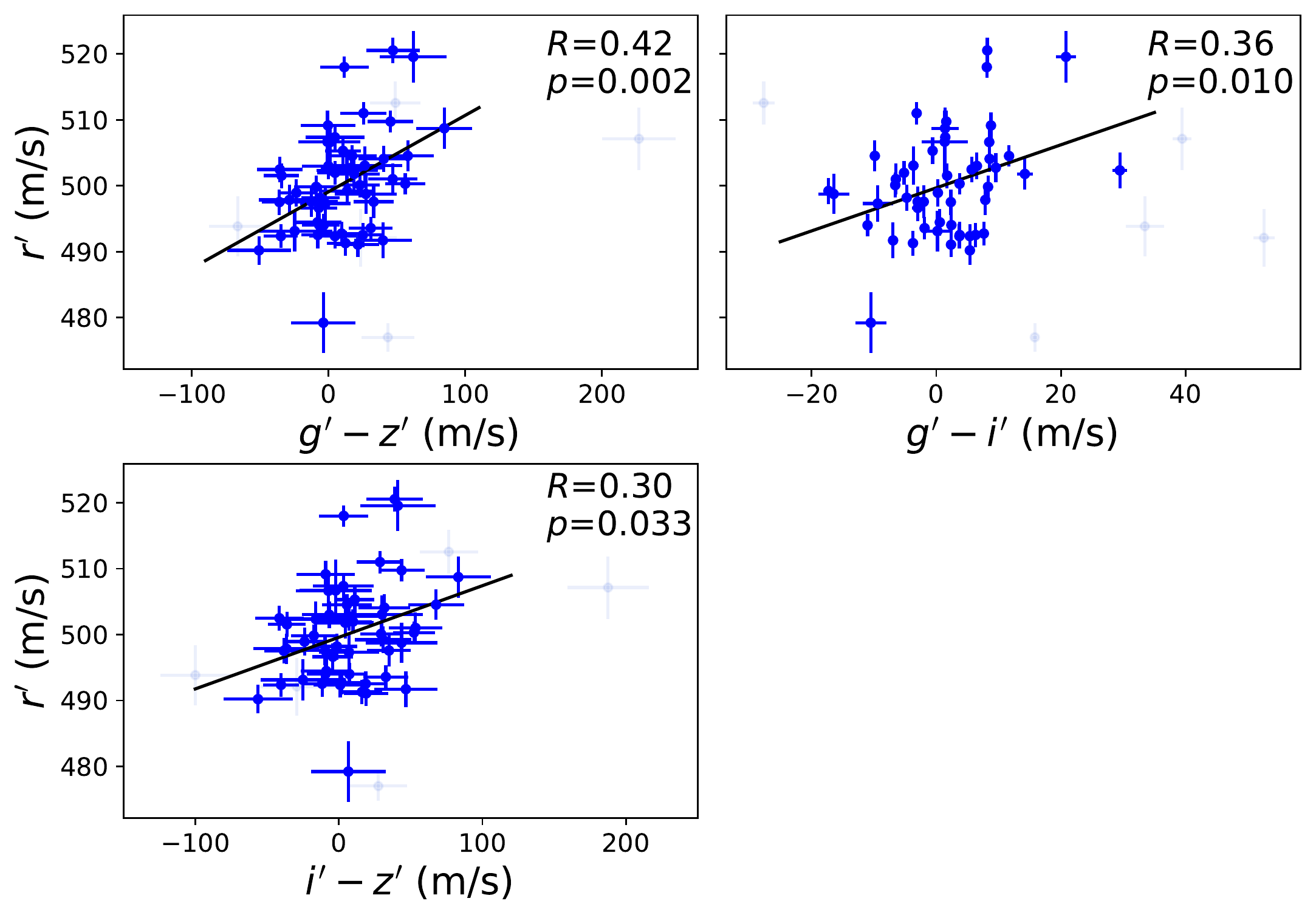}
    \caption{Colour-correlation plots showing the $r^\prime$ band radial velocity as a function of colours calculated from the three remaining bands. The best fitting slope is overplotted in black, and the Pearson R correlation coefficient, and corresponding p-value are included in the upper right corner of each plot. The faded blue points are nights which were sigma-clipped.   
    \label{fig:CARMENES correlations}}
\end{figure}

We first investigated whether chromatic correlations were present in the $r^\prime$ band. In general, one would expect the RV value measured to be independent of the wavelength at which the measurement was taken. That is to say, the RV value in a band should not be correlated with the value in another band, or indeed the difference in RV values between two bands (i.e., an ``RV colour''). We searched for chromatic correlations by comparing the RV value in the $r^\prime$ band with the RV colours calculated from the three remaining bands ($g^\prime-z^\prime$, $g^\prime-i^\prime$, and $i^\prime-z^\prime$). The results are shown in Fig.~\ref{fig:CARMENES correlations}.

In the initial sigma clip that was performed on the CARMENES data set, we removed eight nights which were 3$\sigma$ outliers in all four bands. However, there were a number of other nights which were $>$3$\sigma$ outliers in one or two bands, but not in the others. We have so far retained these nights, however, these can begin to bias our results when analyzing and comparing individual bands. Therefore, at this point if a night is an outlier in any of the four bands, we clip it from out analysis --- another five nights are removed in this way.

For each comparison, we calculate the Pearson correlation coefficient ($R$), as well as the corresponding $p$-value using the \texttt{personr} routine of the \texttt{scipy.stats} package. We additionally fit a linear slope to each case. The best fitting slope and correlation coefficients are also shown in each panel of Fig.~\ref{fig:CARMENES correlations}. As evidenced by the $R$, and $p$-values, the $r^\prime$ band displays significant correlations with each of the three RV colours. \citet{jeffers_carmenes_2022} recently demonstrated that more active host stars have more apparent RV correlations --- K2-18 is known to be moderately active. However, we do not find any correlations from a similar analysis on our reprocessed HARPS dataset.  

Although one would hope for these chromatic correlations to not be present in the first place, it is possible to mitigate their effects via post-processing --- namely we can divide out the best fitting slope in each case to attempt to remove the correlation. We therefore detrend the $r^\prime$ band RVs against all three RV colours, effectively removing all correlations. Re-applying the D'Agostino K-squared test at this juncture results in a $p$-value of $\sim$0.1 --- still a larger deviation from Gaussian that we see for HARPS, but nevertheless a great improvement from the starting value of ${\sim}10^{-9}$. We note that some correlations are also present in the other three bands, however they are not as strong, nor as significant as for the $r^\prime$ band. Given how weakly each of the other three bands contributes to the total RV signal, we chose to only detrend the $r^\prime$ band, and leave the other bands as they were. However, if we do detrend the other three bands as well, it makes no quantitative difference to the results. After the detrending, we then recombine the $r^\prime$ band with the other three bands via a weighted average as before.

\begin{figure}
	\centering
	\includegraphics[width=\columnwidth]{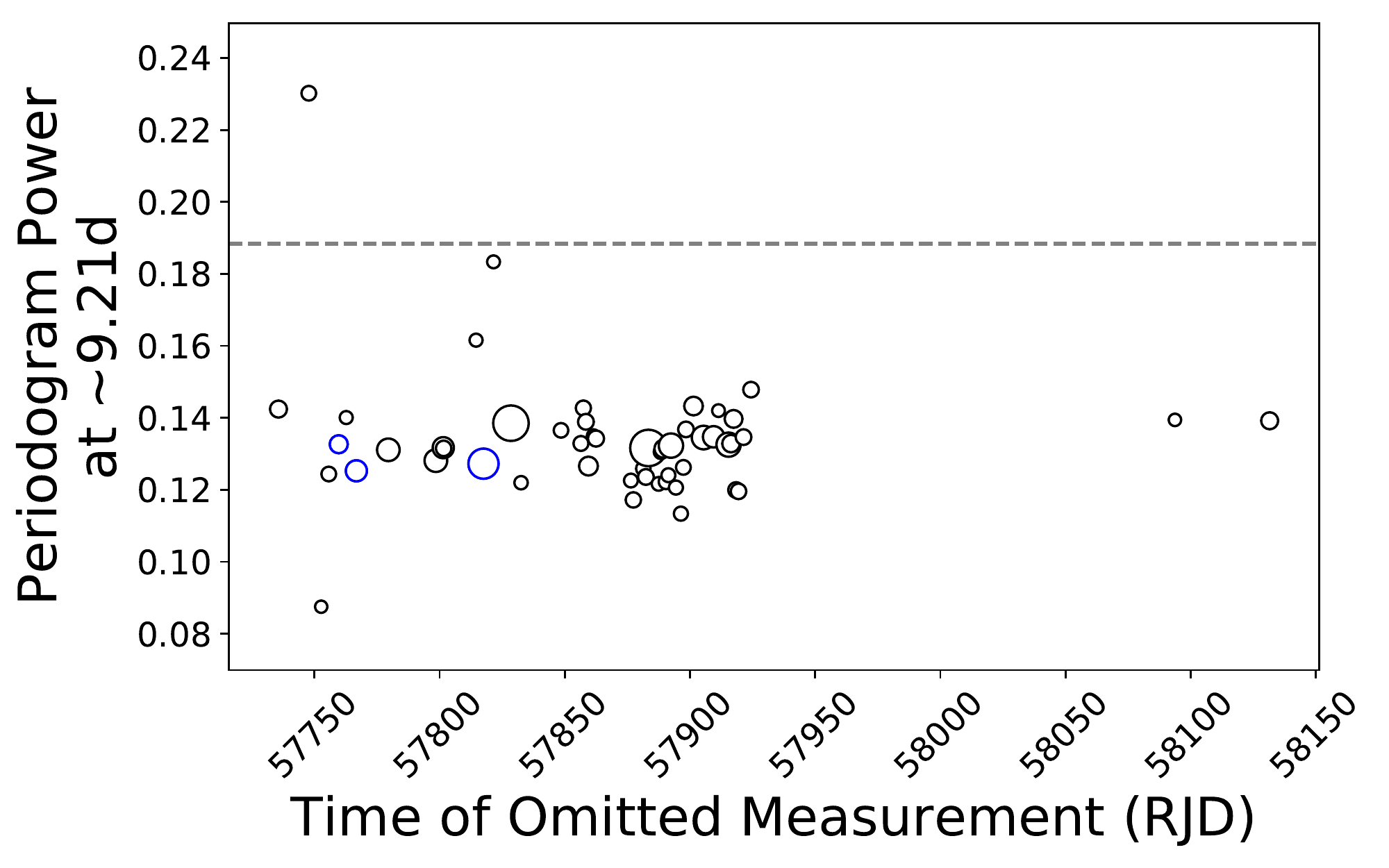}
    \caption{Results of the leave-one-out cross-validation performed on the detrended CARMENES RV time series. The size of the point reflects the relative size of the RV error bars. The dashed grey line denotes the 3$\sigma$ dispersion of the calculated powers. Only one night lies well outside the 3$\sigma$ limit. We do not find the three anomalous nights identified by \citetalias{cloutier_confirmation_2019} (denoted in blue) to suppress the signal of K2-18\,c.
    \label{fig:CARMENES CV}}
\end{figure}

Following \citetalias{cloutier_confirmation_2019} we also performed a leave-one-out cross-validation analysis --- which consists of omitting a single night from the dataset and calculating the power at the period of K2-18\,c (9.207\,d) in a Lomb-Scargle periodogram. In this way, we can understand the impact of individual data points on the ensuing analysis. When analyzing the RV time series presented by \citetalias{sarkis_carmenes_2018}, \citetalias{cloutier_confirmation_2019} hypothesized that there may be a small number of non-outlier nights which are suppressing the signal of K2-18\,c. Indeed, through their leave-one-out cross-validation, they identify three nights, which when removed greatly amplify the signal of K2-18\,c in a Lomb-Scargle periodogram. The results of our leave-one-out cross-validation are shown in Fig.~\ref{fig:CARMENES CV}.

We identify one night (night 1 in our zero-indexed time series, RJD=57747.73) whose removal greatly amplifies the signal of K2-18\,c in the periodogram. Interestingly, this is not one of the nights flagged and removed by \citetalias{cloutier_confirmation_2019} in their analysis. Indeed, none of the three nights identified by \citetalias{cloutier_confirmation_2019} are found to suppress the signal of K2-18\,c (e.g., Fig.~\ref{fig:CARMENES CV}). To attempt to better understand these four nights (the three identified by \citetalias{cloutier_confirmation_2019}, and the one flagged here), and why they may have an outsized impact on the signal of K2-18\,c, we first investigate each night in the context of the observation parameters (e.g., airmass, exposure time, etc.) --- perhaps unfavourable observing conditions render the data taken on these nights to be unreliable. However, by all metrics these four nights appear to have had favourable observing conditions and do not stand out in any way. We then place the nights in the context of our four CARMENES bands to verify if they represent outliers that were somehow missed in our previous analysis. We find that the night flagged during our cross-validation is indeed a slight outlier ($\sim$2.8$\sigma$) in the $r^\prime$ band --- not enough to be captured our previous sigma clipping, but potentially enough to have an impact on the signal of K2-18\,c, especially given the large fractional contribution of the $r^\prime$ band to the final RV measurements (Table~\ref{tab: Band Comparison}). There is though, nothing to suggest the three nights flagged by \citetalias{cloutier_confirmation_2019} to be unreliable. Indeed, their removal or inclusion has minimal impact on the final results. We therefore retain them in all subsequent analyses, but discard the single night flagged by our cross-validation.

\begin{figure}
	\centering
	\includegraphics[width=\columnwidth]{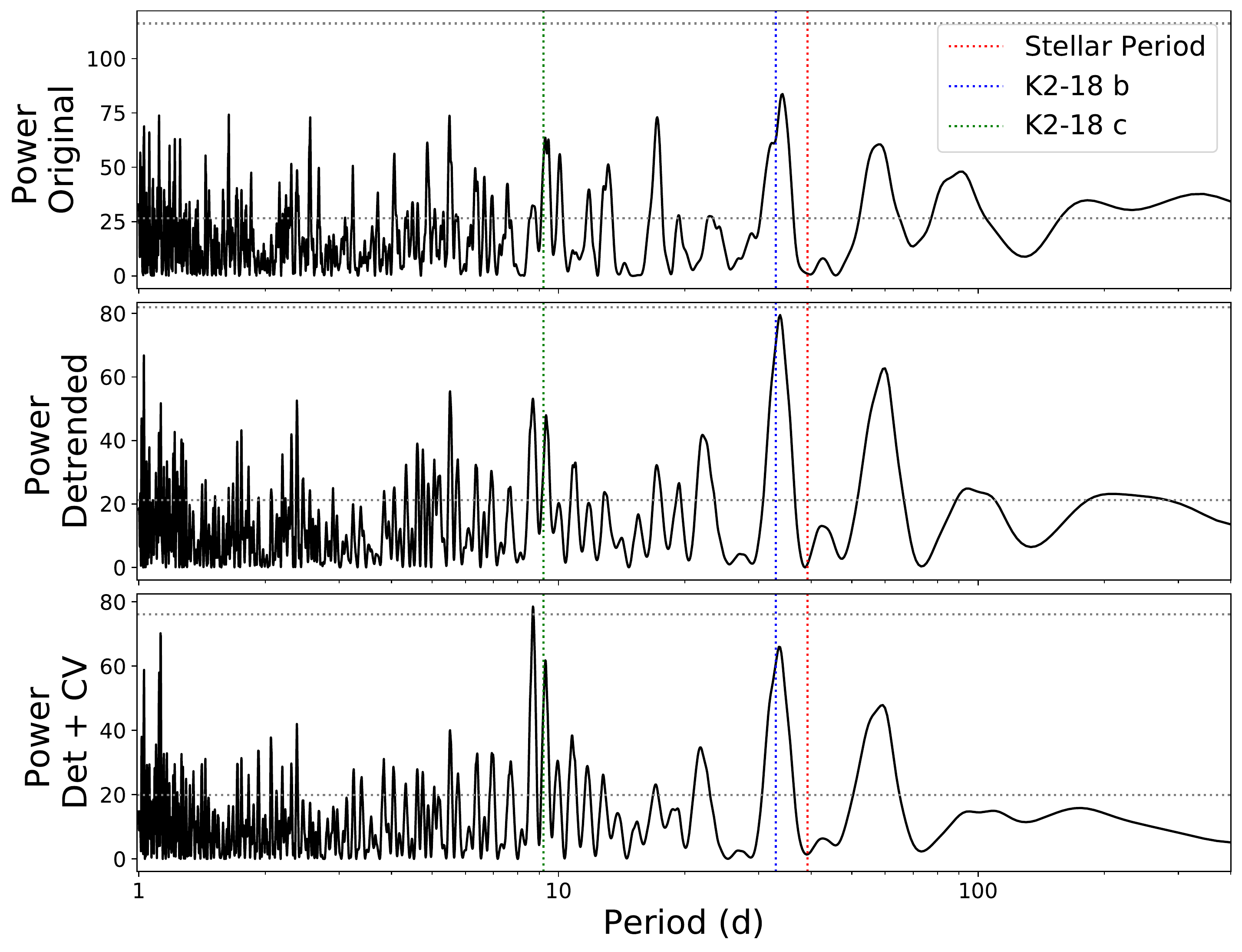}
    \caption{Comparison between the periodogram of the initial CARMENES RV time series (\emph{top}), with the periodogram after detrending (\emph{middle}), and after detrending as well as the removal of the night found to suppress the power of K2-18\,c (\emph{bottom}). The ``noise'' at short $<$5\,d periods is much reduced, and the signals of both K2-18\,b, and c are amplified. Indeed the K2-18\,c signal is even above the 0.1\% FAP level for the first time after our post-processing.
    \label{fig:CARMENES compare}}
\end{figure}

\begin{figure*}
	\centering
	\includegraphics[width=\textwidth]{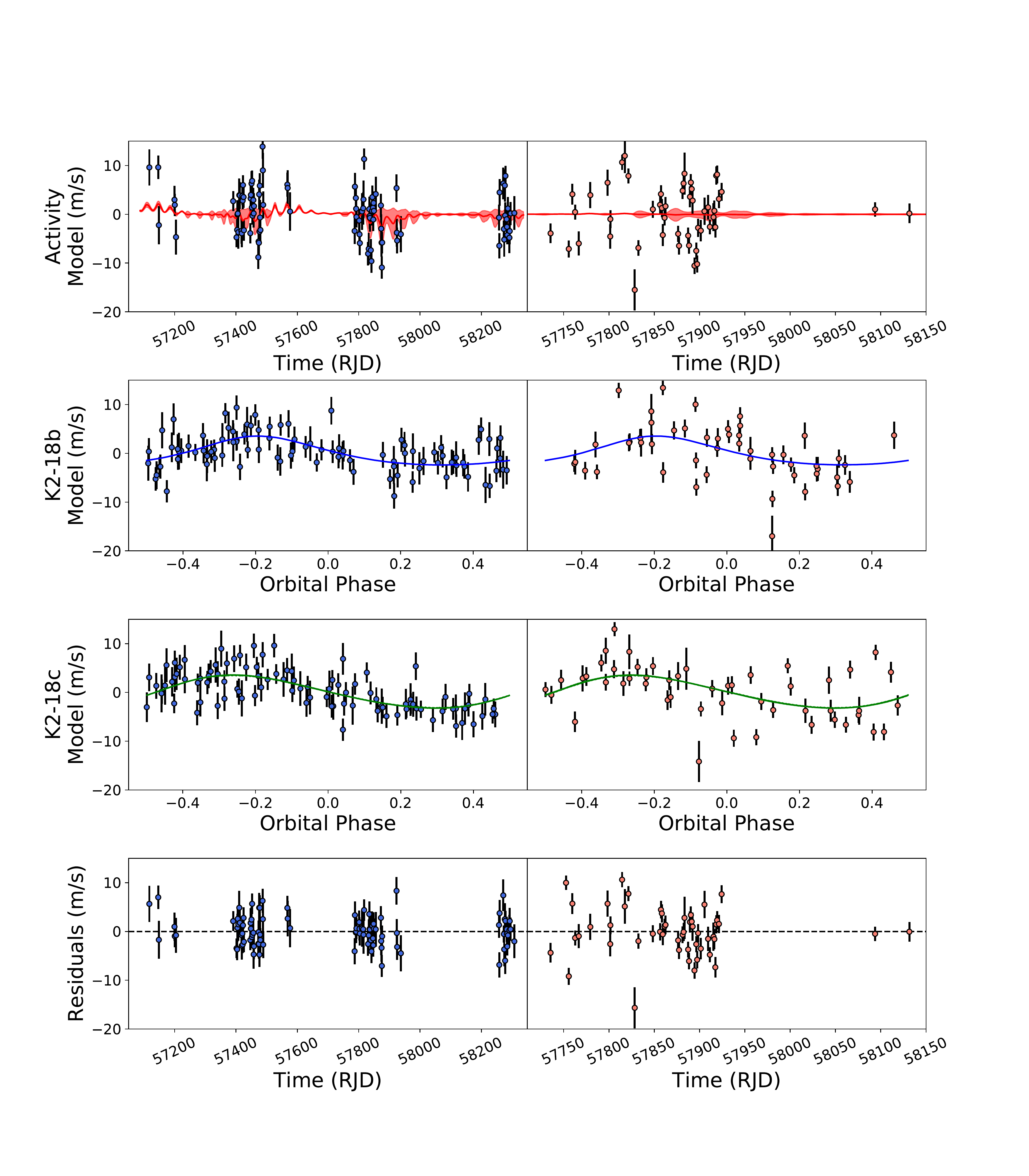}
    \caption{Results of the joint RV/photometry fitting for the HARPS (\emph{left}) and CARMENES (\emph{right}) datasets.  
    \emph{First row}: RV time series with the best fitting GP activity model and 1$\sigma$ envelope overplotted in red. 
    \emph{Second row}: RV time series phase folded to the best fitting period of K2-18\,b. The Keplerian solution is overplotted in blue.
    \emph{Third row}: Same as the second row, but for K2-18\,c, with the Keplerian solution in green.
    \emph{Fourth row}: Residuals to the RV fits.
    \label{fig:JOINT RVModel}}
\end{figure*}

To assess the results of our detrending and cross-validation, we compare the periodograms of the CARMENES RV time series before and after these steps in Fig.~\ref{fig:CARMENES compare}. Indeed, the signal of K2-18\,c is greatly amplified compared to the original time series, and is now above the 0.1\% FAP level. The noise at short periods is also reduced. At this point, we then once again launch a \texttt{juliet} fit of the CARMENES RV data and K2 photometry. The resulting RV models are shown on the right side of Fig.~\ref{fig:JOINT RVModel}. Again, the posteriors are shown in red in Fig.~\ref{fig:JOINT Corner}, and listed in Table~\ref{tab: JOINT Parameters}.

The marginalized posteriors are generally consistent with the parameters derived solely from the HARPS analysis, although often slightly less constraining. K2-18\,b is detected at $>$3$\sigma$ significance, and K2-18\,c at $\sim$2.5$\sigma$. It is notable though that the retrieved RV semi-amplitude of 3.51\,m/s is significantly larger than the $\sim$2.3\,m/s retrieved by \citetalias{cloutier_confirmation_2019} with the CARMENES dataset, and is 1$\sigma$ consistent with the HARPS semi-amplitude. Although \citetalias{cloutier_confirmation_2019} managed to retrieve the signal of K2-18\,c after removing the three nights identified via their leave-one-out cross-validation, the RV semi-amplitude derived from CARMENES was significantly smaller than that derived from HARPS (although they were marginally 1$\sigma$ consistent, mostly due to the extended CARMENES posterior (see their Fig.~6). However, Fig.~\ref{fig:JOINT Corner} shows that in our reanalysis, the CARMENES and HARPS posteriors prefer the same RV semi-amplitude for K2-18\,c. It is therefore possible that although the removal of night flagged through the leave-one-out cross-validation undoes the artificial suppression of the K2-18\,c signal, the uncorrected chromatic trends bias the RV semi-amplitude to lower values. With the chromatic trends corrected, we find much better agreement between the HARPS and CARMENES analyses. Moreover, after our post-processing, the evidence for the presence of K2-18\,c increases from $\Delta\ln Z=-0.41$ to $\Delta\ln Z=2.09$ --- that is to say from ``no evidence'', to a ``moderate detection'' \citep{benneke_how_2013}. 

\subsection{Joint Reanalysis}
\label{sec: Joint Reanalysis}
With the detrended CARMENES dataset in hand, we finally proceed to complete a joint analysis with the HARPS RV, and K2 photometric time series to refine the parameters of the K2-18 system. We once again launch a \texttt{juliet} fit on all datasets together. The assumed priors are once again listed in Table~\ref{tab: JOINT Parameters}. The only differences from the fits on individual instruments is that the GP $\alpha$, $\Gamma$, and $P_{\rm rot}$ hyper-parameters are shared between all instruments (HARPS, CARMENES, and K2), whereas we fit individual amplitudes ($\sigma$) for each instrument. Our fit has a total of 21 free parameters, and the final posterior distributions are shown in green in Fig.~\ref{fig:JOINT Corner}, and listed in Table~\ref{tab: JOINT Parameters}.

\section{Discussion and Conclusions}
\label{sec: Discussion}
We have applied the new LBL method for pRV extraction to two archival datasets of the K2-18 system using the HARPS and CARMENES instruments. Previous analyses, particularly of the CARMENES dataset by \citetalias{sarkis_carmenes_2018} cast doubt on the existence of K2-18\,c, but \citetalias{cloutier_confirmation_2019} showed that the non-detection of the second planet was likely due to a number of anomalous measurements which artificially damped its signal. With our LBL re-reduction, we confirm this hypothesis, and robustly detect K2-18\,c in both datasets. 

The ability of the LBL to subdivide an instrument's full wavelength range into smaller bands when calculating radial velocities has proved to be a powerful tool to enable a deeper understanding of potentially discrepant, or problematic datasets. With the HARPS dataset, it allowed us to verify the self-consistency of extracted RV measurements through intercomparisons of the three different bands. We showed the Gaussian nature of the residuals in each band, thereby confirming the optimal functioning of the \citet{bouchy_fundamental_2001} framework. However, in the case of the CARMENES dataset, the LBL method uncovered concerning substructure, particularly in the $r^\prime$ band, where the dispersion was highly non-Gaussian. Since the $r^\prime$ band contributes most strongly to the final RV measurement than the other three bands we considered (see Fig.~\ref{fig:CARMENES_periodogram}), its anomalies out have an oversized impact on the final RV analysis. It is unclear exactly what caused the highly non-Gaussian nature of the $r^\prime$ band residuals --- especially considering the fact that the other three bands were comparatively much more well-behaved (e.g., Fig.~\ref{fig:CARMENES_hist}). In addition, we found chromatic correlations between the $r^\prime$ band RVs and colours constructed from
the other three CARMENES bands. The correlations were not extremely strong, however they were found to be statistically significant. We detrended the $r^\prime$ band RV measurements in order to remove these chromatic correlations, and in doing so eliminated many of its $>$5$\sigma$ outliers in Fig.~\ref{fig:CARMENES_hist}. 

After this detrending, and the removal of a single night found to be suppressing the signal of K2-18\,c in the CARMENES dataset, the CARMENES periodogram is much cleaner and more closely resembles that of the HARPS data (Fig.~\ref{fig:CARMENES compare}). The signals of both K2-18\,b and c are strong (although not as significant as in the HARPS periodogram), and the amplitudes of spurious signals at short periods are reduced. The comparatively noisier nature of the CARMENES periodogram as opposed to the HARPS is likely due, in large part, to the differing number of nights used in each analysis. Indeed, our HARPS dataset consists of 97 nights, and the CARMENES of only 50 --- it is therefore unsurprising that the signals of interest would be more significant with HARPS than with CARMENES.  

We do remove 14 nights from the CARMENES dataset which are either flagged as $>$3$\sigma$ outliers in at least one band (13 nights), or through the leave-one-out cross-validation (one night). On the face, this may seem like a relatively large fraction of the total RV measurements, but we note that even if we trim fewer nights, for example retain the five nights that were flagged as outliers in our colour correlation analysis, our results remain the same, albeit with less well constrained posteriors for most of our model parameters. 

Another interesting outcome of our analysis of the CARMENES dataset is that the RV semi-amplitude of K2-18\,c is in much better agreement with that derived form the HARPS dataset than was found by \citetalias{cloutier_confirmation_2019}. As can be seen in Fig.~\ref{fig:JOINT Corner}, the $K_c$ posteriors for both instruments agree much better than was found by \citetalias{cloutier_confirmation_2019} (c.f., their Fig.~6). This results in a revision of the minimum mass of K2-18\,c to $6.92 \pm 1.98 \,M_\oplus$ from the published $5.52 \pm 0.84\,M_\oplus$. Our derived mass for K2-18\,b is slightly higher than the value obtained by \citetalias{cloutier_confirmation_2019}, but still consistent at the 1$\sigma$ level. 

Our work also demonstrates the power of Nested Sampling algorithms for retrieval analyses. \citetalias{cloutier_confirmation_2019} misidentified the period of K2-18\,c as their MCMC algorithm could not adequately capture the multi-modal nature of its posterior probability distribution. However, our Nested Sampling algorithm does not suffer from the same limitations. Furthermore, we obtain the Bayesian evidence ``for free'', allowing model comparison between orbital solutions with a 8.997\,day and 9.207\,day period which robustly confirms our findings.  

\section*{Acknowledgements}
MR would like to acknowledge funding from the National Sciences and Research Council of Canada (NSERC), the Fonds de Recherche du Qu\'ebec - Nature et Technologies (FRQNT), and the Institut de Recherche sur les Exoplanètes (iREx) for support towards his doctoral studies. MR would also like to thank Thomas Vandal and Néstor Espinoza for helpful discussion about RV model fitting and GP processes. 
PJA, JAC, and IR acknowledge financial support from the Spanish Agencia Estatal de Investigaci\'on of the Ministerio de Ciencia e Innovaci\'on (AEI-MCIN) and the European FEDER/ERF funds through projects
 PGC2018-098153-B-C33,     
 PID2019-109522GB-C51/52/53/54,     
and the Centre of Excellence ``Severo Ochoa'' and ``Mar\'ia de Maeztu'' awards to the Instituto de Astrof\'isica de Andaluc\'ia (SEV-2017-0709), Centro
de Astrobiolog\'ia (MDM-2017-0737), and the Institut de Ci\`encies de l'Espai (CEX2020-001058-M).

\section*{Software}
\begin{itemize}
    \renewcommand\labelitemi{--}
    \item \texttt{astropy}; \citet{astropy:2013, astropy:2018}
    \item \texttt{ipython}; \citet{PER-GRA:2007}
    \item \texttt{juliet}; \citet{espinoza_juliet_2019}
    \item \texttt{matplotlib}; \citet{Hunter:2007}
    \item \texttt{numpy}; \citet{harris2020array}
    \item \texttt{pymultinest}; \citet{buchner_statistical_2016}
    \item \texttt{scipy}; \citet{2020SciPy-NMeth}
\end{itemize}

\section*{Data Availability}
The full LBL RV time series can be downloaded from the CDS. The HARPS spectra are publicly available from the ESO science archive.

\bibliography{K218.bib}
\bibliographystyle{aasjournal}

\appendix
\section{Additional Figures and Tables}

\begin{figure*}
	\centering
	\includegraphics[width=\textwidth]{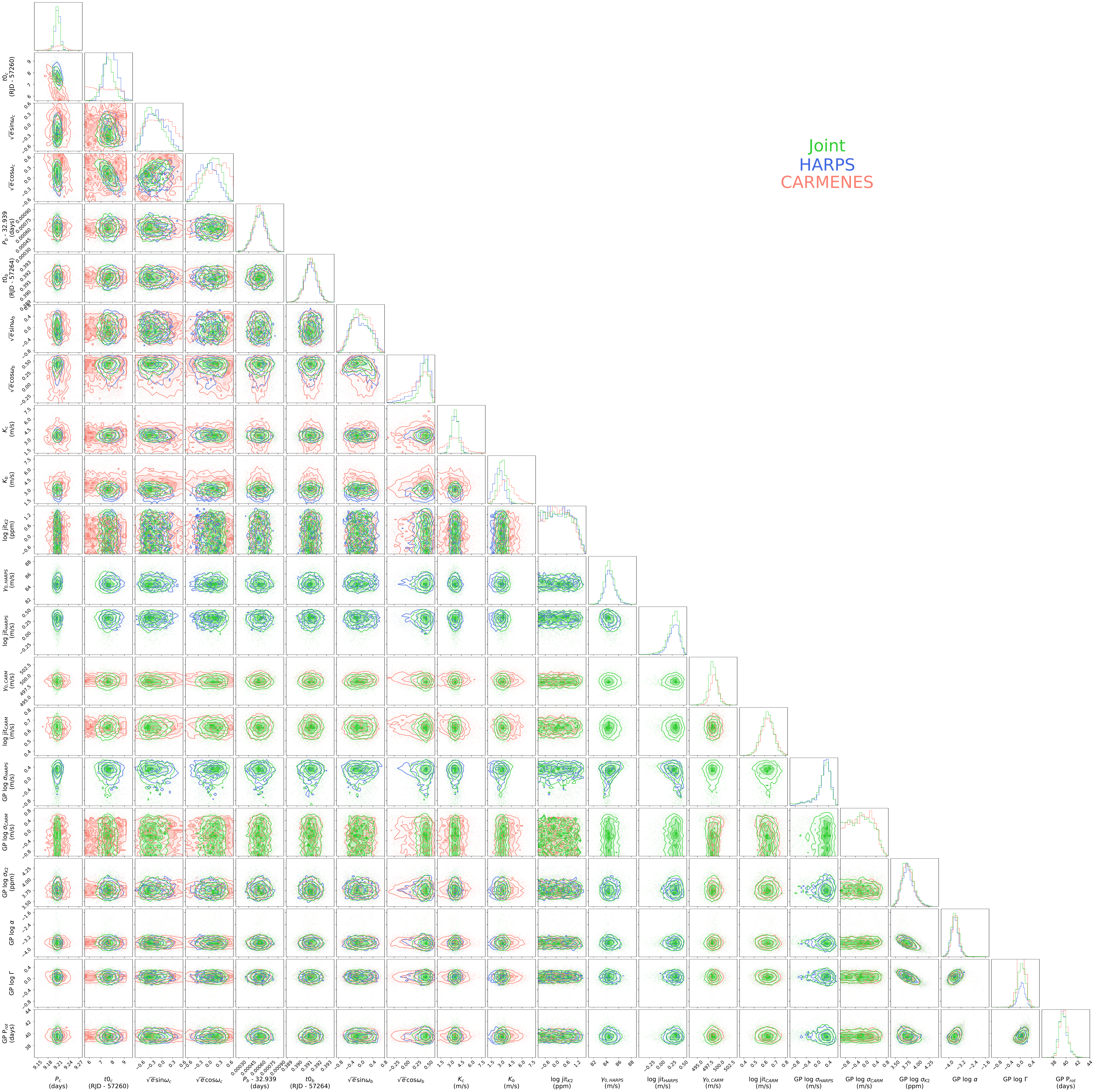}
    \caption{Posterior distributions for all fitted parameters for the joint HARPS + CARMENES + K2 (green) fit, as well as for the individual HARPS + K2 (blue) and CARMENES + K2 (red) fits. Labels capping each column are the median, and 1$\sigma$ errors for the joint fit. In general, the HARPS and CARMENES posteriors are nearly identical, with CARMENES proving to be 1$\sigma$ consistent, although marginally less constraining for the RV semi-amplitudes of both planets, as well as the orbital parameters of K2-18\,c.
    \label{fig:JOINT Corner}}
\end{figure*}

\begin{landscape}
\begin{table}
\caption{Fitted Model Parameters and their Prior Distributions}
\label{tab: JOINT Parameters}
    \begin{threeparttable}
        \begin{tabular}{ccccc}
            \hline
            \hline
            Model Parameter & Prior Distribution & \multicolumn{3}{c}{Posterior Median with 16th and 84th Percentiles} \\
             &  & \textit{HARPS + K2} & \textit{CARMENES + K2} & \textit{Joint} \\
            \hline
            
            \textit{System Parameters} & & & & \\
            Systemic Velocity (HARPS), $\gamma_{0, HARPS}$ [m/s] & $\mathcal{U}(-1000, 1000)$ & $84.57^{+0.82}_{-0.67}$ & -- & $85.50^{+0.88}_{-0.70}$ \\
            \vspace{1mm}Systemic Velocity (CARMENES), $\gamma_{0, CARM}$ [m/s] & $\mathcal{U}(-1000, 1000)$ & -- & $498.92^{+0.97}_{-0.88}$ & $498.56^{+0.93}_{-0.87}$ \\
            
            \textit{GP Hyperparameters} & & & & \\
            GP Amplitude (HARPS), $\sigma_{HARPS}$ [m/s] & $\mathcal{J}(0.1, 100)$ & $1.933^{+0.860}_{-0.911}$ & -- & $1.818^{+0.892}_{-0.944}$ \\
            GP Amplitude (CARMENES), $\sigma_{CARM}$ [m/s] & $\mathcal{J}(0.1, 100)$ & -- & $0.679^{+1.375}_{-0.473}$ & $0.617^{+1.310}_{-0.435}$ \\
            GP Amplitude (K2), $\sigma_{K2}$ [ppm] & $\mathcal{J}(10^{-3}, 10^{6})$ & $5871.100^{+2265.863}_{-1434.768}$ & $5917.908^{+2000.039}_{-1355.728}$ & $5874.013^{+2313.581}_{-1398.826}$ \\
            GP log Exponential Timescale, $\alpha$ [days] & $\mathcal{U}(-5, 5)$ & $-3.612^{+0.229}_{-0.229}$ & $-3.596^{+0.231}_{-0.212}$ & $-3.594^{+0.251}_{-0.235}$ \\
            GP log Coherence, $\Gamma$ & $\mathcal{U}(-5, 5)$ & $0.052^{+0.118}_{-0.120}$ & $0.045^{+0.118}_{-0.116}$ & $0.052^{+0.129}_{-0.131}$ \\
            GP Periodic Timescale, $P_{rot}$ [days] & $\mathcal{U}(0, 100)$ & $39.522^{+0.740}_{-0.613}$ & $39.649^{+0.671}_{-0.568}$ & $39.556^{+0.740}_{-0.626}$ \\
            Additive Jitter (HARPS), $jit_{HARPS}$ [m/s] & $\mathcal{J}(10^{-2,} 10^{2})$ & $1.961^{+0.522}_{-0.488}$ & -- & $1.957^{+0.500}_{-0.545}$ \\
            Additive Jitter (CARMENES), $jit_{CARM}$ [m/s] & $\mathcal{J}(10^{-2}, 10^{2})$ & -- & $4.364^{+0.644}_{-0.563}$ & $4.234^{+0.582}_{-0.546}$ \\
            \vspace{1mm}Additive Jitter (K2), $jit_{K2}$ [ppm] & $\mathcal{J}(10^{-1}, 10^{4})$ & $1.497^{+7.426}_{-1.252}$ & $1.361^{+5.804}_{-1.085}$ & $1.387^{+7.536}_{-1.160}$ \\

            \textit{K2-18\,b} & & & & \\
            Period, $P_b$ [days] & $\mathcal{N}(32.93962, 10^{-4})^\dagger$ & $32.9396\pm10^{-4}$ & $32.9396\pm10^{-4}$ & $32.9396\pm10^{-4}$\\
            Time of Inferior Conjunction, $T_{0, b}$ [RJD] & $\mathcal{N}(57264.39142, 6.4\times10^{-4})^\dagger$ & $57264.3914\pm0.0007$ & $57264.3914\pm0.0007$ & $57264.3914\pm0.0007$ \\
            RV Semi-Amplitude, $K_b$ [m/s] & $mod\mathcal{J}(1, 20)$ & $2.628^{+0.678}_{-0.736}$ & $3.699^{+1.207}_{-1.183}$ & $3.112^{+0.557}_{-0.564}$ \\
            $\sqrt{e_b}\sin\omega_b$ & $\mathcal{U}(-1, 1)$ & $-0.059^{+0.396}_{-0.311}$ & $-0.024^{+0.332}_{-0.342}$ & $-0.062^{+0.313}_{-0.272}$ \\
            $\sqrt{e_b}\cos\omega_b$ & $\mathcal{U}(-1, 1)$ & $0.393^{+0.104}_{-0.199}$ & $0.331^{+0.167}_{-0.320}$ &$0.408^{+0.078}_{-0.124}$ \\            
            \vspace{1mm}Planet Mass$^*$, $M_b$ [$M_\oplus$] & & $7.839^{+1.881}_{-2.327}$ & $10.902^{+3.742}_{-3.658}$ & $9.510^{+1.567}_{-1.890}$ \\
  
            \textit{K2-18\,c} & & & & \\
            Period, $P_c$ [days] & $\mathcal{U}(8, 10)$ & $9.208^{+0.007}_{-0.007}$ & $9.343^{+0.601}_{-0.160}$ & $9.207^{+0.007}_{-0.006}$ \\
            Time of Inferior Conjunction, $T_{0, c}$ [RJD] & $\mathcal{U}(57262, 57272)$ & $57267.831^{+0.536}_{-0.462}$ & $57266.847^{+1.370}_{-1.241}$ & $57267.581^{+0.480}_{-0.467}$ \\
            RV Semi-Amplitude, $K_c$ [m/s] & $mod\mathcal{J}(1, 20)$ & $3.620^{+0.521}_{-0.531}$ & $3.467^{+1.172}_{-1.409}$ & $3.568^{+0.439}_{-0.457}$ \\
            $\sqrt{e_c}\sin\omega_c$ & $\mathcal{U}(-1, 1)$ & $-0.166^{+0.302}_{-0.265}$ & $-0.023^{+0.454}_{-0.449}$ &$-0.251^{+0.286}_{-0.208}$ \\
            $\sqrt{e_c}\cos\omega_c$ & $\mathcal{U}(-1, 1)$ & $-0.013^{+0.263}_{-0.261}$ & $0.059^{+0.380}_{-0.435}$ & $0.114^{+0.224}_{-0.257}$ \\
            \vspace{1mm}Minimum Planet Mass$^*$, $M_c\sin i_c$ [$M_\oplus$] & & $7.022^{+1.141}_{-1.151}$ & $6.372^{+2.507}_{-2.832}$ & $6.922^{+0.962}_{-0.991}$ \\
            \hline
            
        \end{tabular}
        \begin{tablenotes}
            \small
            \item \textbf{Notes}: $^*$ Denotes a derived parameter. Masses were calculated assuming a stellar mass of $0.4951\pm0.0043\,M_\odot$ \citep{benneke_water_2019}, and an inclination for K2-18\,b of $89.5785^\circ$ \citep{benneke_spitzer_2017}. $^\dagger$ Based on the transit measurements of \citet{benneke_spitzer_2017}.  -- Indicates that a parameter was not included in that fit.  $\mathcal{U}$ represents a uniform prior with equal probability per unit. $\mathcal{J}$ represents a Jeffreys prior with equal probability per decade. $\mathcal{N}(x, y)$ represents a normally distributed prior centered at $x$, with a width of $y$. $mod\mathcal{J}(x, y)$ represents a modified Jefrrey's prior, which behaves like a uniform prior for values $<x$, and a Jeffrey's prior $>x$. 
        \end{tablenotes}
    \end{threeparttable}
\end{table}
\end{landscape}

\begin{landscape}
\begin{table}
\caption{Full HARPS-LBL Time Series }
\label{tab: HARPS data}
    \begin{threeparttable}
        \begin{tabular}{cccccccccc}
            \hline
            \hline
            BJD - 2400000 & RV & $\sigma$RV & RV ($u^\prime$) & $\sigma$RV ($u^\prime$) & RV ($g^\prime$) & $\sigma$RV ($g^\prime$) & RV ($r^\prime$) & $\sigma$RV ($r^\prime$) & dLW \\
            & (m/s) & (m/s) & (m/s) & (m/s) & (m/s) & (m/s) & (m/s) & (m/s) & \\
            \hline
            
            57117.56587 & 94.124 & 3.696 & 172.796 & 84.828 & 100.031 & 7.318 & 91.456 & 5.174 & 114401.696\\
            57146.52695 & 94.141 & 2.408 & 132.574 & 79.442 & 92.804 & 4.634 & 92.853 & 3.414 & 172287.280\\
            57146.64607 & 101.863 & 3.383 & 10.130 & 83.555 & 94.013 & 6.779 & 105.664 & 4.693 & 124530.558\\ 
            57148.51885 & 82.284 & 3.904 & -1.255 & 81.273 & 84.830 & 7.928 & 82.853 & 5.387 & 125182.600\\ 
            57199.50391 & 87.493 & 2.876 & 135.617 & 84.608 & 92.620 & 5.164 & 84.240 & 4.275 & 261253.532\\ 
            57200.50311 & 86.428 & 2.292 & -53.884 & 77.141 & 84.556 & 4.219 & 86.720 & 3.348 & 259047.599\\ 
            57204.49117 & 79.839 & 3.592 & 139.444 & 85.363 & 86.182 & 6.874 & 79.002 & 5.128 & 246499.966\\ 
            57390.84508 & 87.206 & 2.057 & 78.025 & 69.617 & 84.391 & 3.401 & 86.682 & 3.324 & 330037.719\\ 
            57401.77922 & 79.783 & 2.011 & 25.376 & 66.801 & 78.612 & 3.413 & 81.240 & 3.138 & 184148.180\\ 
            57403.82687 & 81.312 & 2.187 & 46.689 & 62.941 & 82.177 & 3.736 & 80.179 & 3.423 & 160552.608\\ 
            \hline
            
        \end{tabular}
        \begin{tablenotes}
            \small
            \item \textbf{Notes}: Only the first 10 rows of this table are shown to demonstrate its format; a machine readable version is available in the online material. The full data set can also be downloaded from the CDS. 
        \end{tablenotes}
    \end{threeparttable}
\end{table}
\end{landscape}

\begin{landscape}
\begin{table}
\caption{Full CARMENES-LBL Time Series }
\label{tab: CARMENES data}
    \begin{threeparttable}
        \begin{tabular}{cccccccccccc}
            \hline
            \hline
            BJD - 2400000 & RV & $\sigma$RV & RV ($g^\prime$) & $\sigma$RV ($g^\prime$) & RV ($r^\prime$) & $\sigma$RV ($r^\prime$) & RV ($i^\prime$) & $\sigma$RV ($i^\prime$) & RV ($z^\prime$) & $\sigma$RV ($z^\prime$)& dLW \\
            & (m/s) & (m/s) & (m/s) & (m/s) & (m/s) & (m/s) & (m/s) & (m/s) & (m/s) & (m/s) &\\
            \hline
            
            57735.61776 & 491.428 & 2.018 & 500.577 & 8.205 & 490.197 & 2.171 & 495.176 & 7.576 & 551.482 & 31.646 & 297762.957\\ 
            57747.73287 & 519.376 & 1.744 & 518.692 & 6.071 & 520.544 & 1.909 & 510.504 & 6.221 & 471.606 & 25.747 & 180535.708\\ 
            57752.68374 & 517.259 & 1.466 & 517.348 & 4.779 & 518.008 & 1.610 & 509.234 & 5.439 & 505.854 & 22.551 & 215226.257\\ 
            57755.70984 & 491.991 & 1.746 & 491.965 & 6.118 & 492.357 & 1.907 & 488.247 & 6.362 & 487.293 & 24.980 & 160751.602\\ 
            57759.69413 & 503.357 & 2.130 & 494.338 & 8.002 & 504.532 & 2.322 & 504.182 & 7.465 & 436.404 & 27.072 & 118527.351\\ 
            57762.68384 & 499.807 & 1.551 & 503.096 & 5.261 & 499.834 & 1.699 & 494.779 & 5.699 & 512.192 & 21.533 & 150647.335\\ 
            57766.73467 & 492.742 & 2.486 & 496.567 & 9.601 & 491.709 & 2.702 & 503.485 & 8.832 & 456.798 & 30.663 & 30416.110\\ 
            57779.49771 & 502.572 & 2.677 & 498.548 & 10.730 & 503.059 & 2.887 & 502.210 & 9.911 & 472.121 & 38.235 & 310571.531\\ 
            57787.47672 & 493.241 & 7.087 & 436.431 & 26.143 & 499.551 & 7.767 & 492.043 & 25.080 & 424.833 & 59.606 & 562879.493\\ 
            57791.46270 & 492.963 & 4.050 & 526.776 & 16.388 & 492.100 & 4.383 & 474.215 & 14.681 & 503.398 & 42.829 & 252616.256\\ 
            \hline
            
        \end{tabular}
        \begin{tablenotes}
            \small
            \item \textbf{Notes}: Only the first 10 rows of this table are shown to demonstrate its format; a machine readable version is available in the online material. The full data set can also be downloaded from the CDS. 
        \end{tablenotes}
    \end{threeparttable}
\end{table}
\end{landscape}



\bsp	
\label{lastpage}
\end{document}